\documentclass[conference]{IEEEtran}
\IEEEoverridecommandlockouts

\usepackage{cite}
\usepackage{amsmath,amssymb,amsfonts,amsthm}
\usepackage{mathtools}
\usepackage{algorithm}
\usepackage{algorithmic}
\usepackage{graphicx}
\usepackage{textcomp}
\usepackage{xcolor}
\usepackage{gensymb}
\usepackage{verbatim}
\def\BibTeX{{\rm B\kern-.05em{\sc i\kern-.025em b}\kern-.08em T\kern-.1667em\lower.7ex\hbox{E}\kern-.125emX}}

\newtheorem{proposition}{Proposition}

\newtheorem{lemma}{Lemma}
\newtheorem{remark}{Remark}

\newtheorem{corollary}{Corollary}

\newcommand{\rmd}{{\,\mathrm{d}}}
\renewcommand{\IEEEQED}{\IEEEQEDopen}

\begin{document}
\bstctlcite{IEEEexample:BSTcontrol}

\title{Performance Analysis of IOS-Assisted NOMA System with Channel Correlation and Phase Errors}
\author{
\thanks{This work was supported by EPSRC grant number EP/T02612X/1. For the  purpose of Open Access, the author has applied a CC BY public copyright licence to any Author Accepted Manuscript (AAM) version arising from this submission. (Corresponding author: Gaojie Chen)}
        Tianxiong Wang,
        Mihai-Alin~Badiu,~\IEEEmembership{Member, IEEE,}
        Gaojie Chen,~\IEEEmembership{Senior Member, IEEE,}
        \\ and Justin~P.~Coon,~\IEEEmembership{Senior Member, IEEE,}
\thanks{T. Wang, M. A. Badiu and J. P. Coon are with the Department of Engineering Science, University of Oxford, Oxford, OX1 3PJ, U.K. (e-mail: \{tianxiong.wang, mihai.badiu, justin.coon\}@eng.ox.ac.uk).}
\thanks{G. Chen is with 5GIC \& 6GIC, Institute for Communication Systems (ICS), University of Surrey, Guildford, GU2 7XH, United Kingdom (e-mail: gaojie.chen@surrey.ac.uk).}
\vspace{-2em}}

\maketitle

\begin{abstract}
In this paper, we investigate the performance of an intelligent omni-surface (IOS) assisted downlink non-orthogonal multiple access (NOMA) network with phase quantization errors and channel estimation errors, where the channels related to the IOS are spatially correlated. First, upper bounds on the average achievable rates of the two users are derived. Then, channel hardening is shown to occur in the proposed system, based on which we derive approximations of the average achievable rates of the two users. The analytical results illustrate that the proposed upper bound and approximation on the average achievable rates are asymptotically equivalent in the number of elements. Furthermore, it is proved that the asymptotic equivalence also holds for the average achievable rates with correlated and uncorrelated channels. { Additionally, we extend the analysis by evaluating the average achievable rates for IOS assisted orthogonal multiple access (OMA) and IOS assisted multi-user NOMA scenarios.} Simulation results corroborate the theoretical analysis and {  demonstrate that: i) low-precision elements with only two-bit phase adjustment can achieve the performance close to the ideal continuous phase shifting scheme; ii) The average achievable rates with correlated channels and uncorrelated channels are asymptotically equivalent in the number of elements; iii) IOS-assisted NOMA does not always perform better than OMA due to the reconfigurability of IOS in different time slots.}

\end{abstract}


\section{Introduction}
The future sixth-generation (6G) is expected to support billions of connected devices with significant demands in spectrum efficiency, reliability, latency, and connectivity \cite{dang2020should}. To handle these challenging requirements, a wide range of technologies, including massive multiple-input multiple-output (MIMO), network densification, and millimeter-wave (mm-wave) communications, have been investigated extensively in recent years \cite{david20186g}. Among various technologies, non-orthogonal multiple access (NOMA) has been regarded as a superior multiple access technology in the future wireless networks due to its promising advantages in terms of spectrum efficiency \cite{tang2019energy,alkhawatrah2019buffer,dai2015non}. For example, multiple users could share the same resource block simultaneously in NOMA \cite{zhu2020power}. As pointed in \cite{wang2021beamforming}, NOMA has better performance than orthogonal multiple access (OMA) when the channels gains of users are remarkably different. Conventionally, the channel gains of the users are determined by the uncontrollable wireless propagation environment \cite{b8, huang2021buffer}. Hence, the applications of NOMA are limited in conventional wireless networks \cite{chen2016application}.

With the rapid development of metasurfaces, reconfigurable intelligent surfaces (RISs) have been recognized as a promising technology in the next-generation wireless communications networks \cite{b4}. Physically, a RIS consists of a large number of passive reflecting elements, each of which can shift the phases of the incident electromagnetic (EM) waves. Through intelligent phase shifting, a RIS could increase or decrease the composite channel gains of different users \cite{huang2021multi, Q2020T}. Due to the advantage of RIS in reconfiguring the channels, many existing works investigated the potential of applying RIS in NOMA systems to broaden the applications of NOMA. {The authors in \cite{ding2020simple} and \cite{yue2021performance} analyzed the performance of RIS assisted NOMA networks where each element of the RIS applies  1-bit coding scheme. The authors in \cite{hou2020reconfigurable} proposed a priority oriented design where the RIS is controlled to boost the channel of the prioritized NOMA user and evaluated the system performance.}
The uplink outage probability of a RIS-aided NOMA network was derived in \cite{tahir2020analysis}. Considering coherent and random phase shifting, the authors in \cite{ding2020impact} examined the impact of these two phase-shifting schemes on the outage performance of a RIS-assisted NOMA system. The capacity of a RIS-empowered NOMA network with multiple users was analyzed in \cite{mu2021capacity}. 
In \cite{zheng2020intelligent}, the passive beamforming at RIS were optimized for minimizing the transmit power and the performance between RIS-assisted NOMA and RIS-assisted OMA were compared.

\begin{figure}[t!]
    \centerline{\includegraphics[scale=0.18]{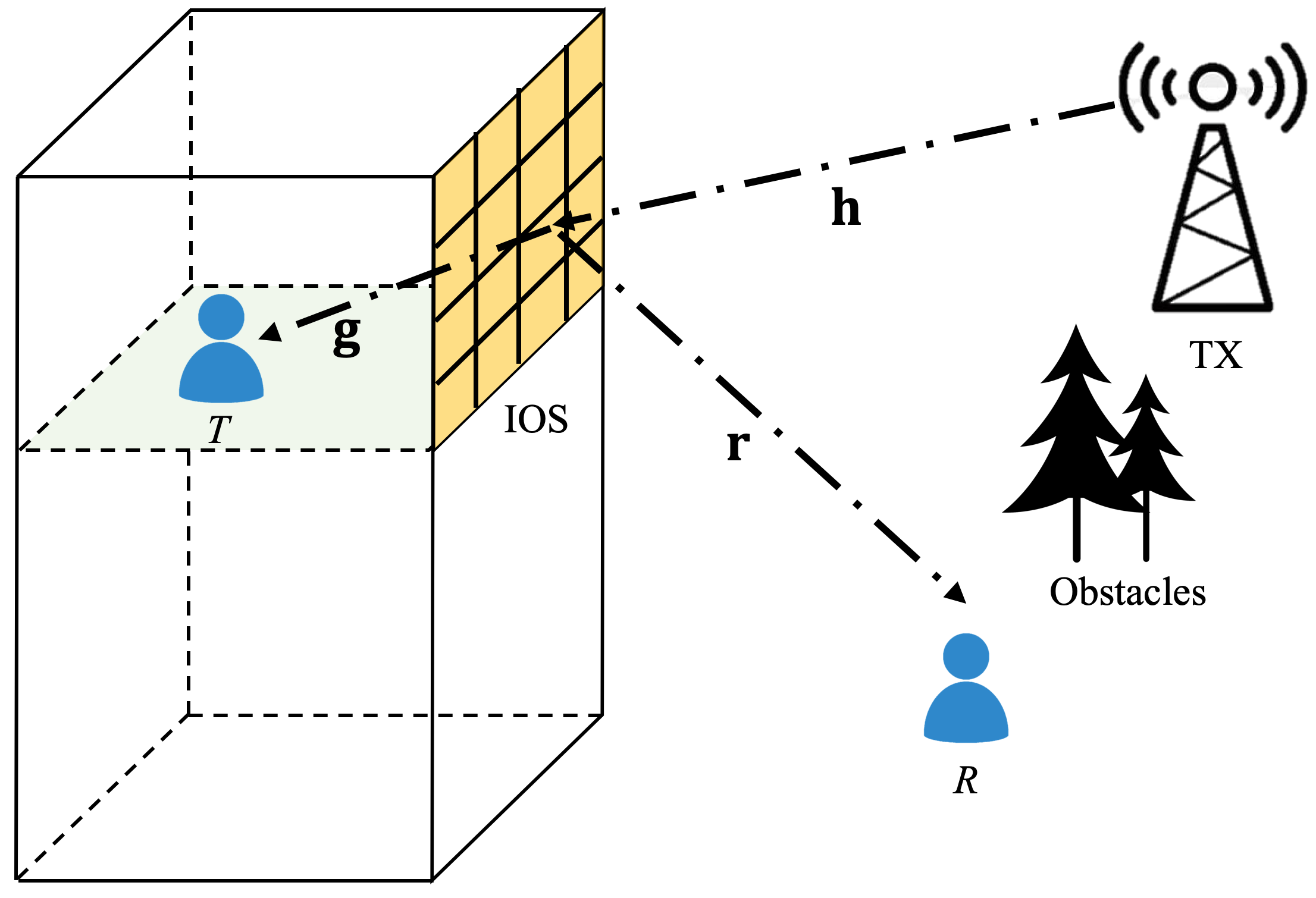}}   
    \caption{System model of the IOS-assisted NOMA system.}
    \label{sys1}
\end{figure}

However, the above works considered the reflecting-only RIS, which requires users located on the same side as the transmitter. As a result, the users located on the other side of the reflecting-only RIS are blocked, which limits the service coverage of RIS. The concept of intelligent omni-surface (IOS) with dual signal refraction and reflection functions was proposed to tackle this problem in some recent works. In contrast to the conventional reflecting-only RIS, signals impinging on one side of the IOS can be reflected and transmitted to users on the same and opposite sides of the surface as the transmitter, as illustrated in Fig.~\ref{sys1}. A prototype of IOS was developed by NTT DOCOMO, Japan \cite{proto2020NTT}. By dynamically adjusting the space between the metasurface and the substrate, the incident radio waves can be entirely reflected, entirely transmitted, or simultaneously reflected and transmitted without attenuation. The authors in \cite{zhang2020beyond} optimized the spectral efficiency of a single-input single-output (SISO) communication link assisted by an IOS where the transmitting and reflecting signals share the same phase shift and showed the deployment of IOS can significantly enlarge the wireless coverage. As a step further, the joint active and passive beamforming design in a downlink IOS-assisted communication network with a multiple-antenna access point (AP) and multiple users was studied in \cite{zeng2021reconfigurable, zhang2021intelligent1}. {  The authors in \cite{zhang2022meta} proposed to use an IOS to alleviate the inter-cell interference between two independent access points (APs).} IOS-empowered physical layer security was investigated in \cite{fang2022intelligent}. The authors in \cite{liu2021star} proposed a new structure of IOS, also referred to as simultaneous transmitting and reflecting reconfigurable intelligent surface (STAR-RIS), where the phase shifts of the reflecting and transmitting signals can be configured independently. The authors in \cite{xu2021star} studied the near-field and far-field channel models of an IOS-assisted communication network. In \cite{9570143}, active beamforming at the base station (BS) and passive transmitting and reflecting beamforming at the IOS were jointly optimized to minimize the power consumption of the network. IOS-aided NOMA systems were also studied in several recent works. Under the constraint on the communication requirement of each user, the coverage range of an IOS-assisted NOMA network was optimized in \cite{wu2021coverage}. The authors in \cite{aldababsa2021simultaneous} proposed a multi-user IOS-assisted NOMA system and optimized the sum-rate with requirements on the quality-of-service of individual users. The outage probability and diversity gain of an IOS empowered NOMA system was derived in \cite{zhang2022star}.

Although IOS-assisted communications have attracted significant research interests recently, the existing works have not covered several aspects of performance analysis and system design. {  First, most of the current works do not consider the effect of channel correlation in the performance analysis of IOS-assisted systems.} However, as pointed in \cite{bjornson2020rayleigh}, the assumption of independent channels is valid only when the half-wavelength spaced elements are placed in a linear array. Hence, the performance of IOS-assisted networks with correlated channels requires more investigation. Second, the existing works on the performance analysis of IOS-assisted networks usually assume the channel estimation and phase adjustment are perfect. However, imprecise channel estimation and phase quantization lead to phase errors in practice. {  It is essential to include phase errors in the performance analysis.}

This paper presents novel results on the average achievable rates of an energy-splitting IOS-assisted NOMA network with correlated channels and imperfect phase adjustment. The main contributions of this paper are summarized as follows: 
\begin{enumerate}
    \item We analyze a downlink energy-splitting IOS-assisted NOMA communication system with spatially correlated channels and phase errors. Upper bounds on the average achievable rates of the users with strong and weak channel conditions are derived by using Jensen's inequality.
    \item We reveal the channel hardening effect in the IOS-assisted NOMA system with correlated channels and phase errors. Then, using the results of the channel hardening effect, two approximate expressions of the average achievable rates are obtained. {  We also analyze the impact of discrete phase shifts on the performance and show that the improvement brought by having more bits for phase adjustment decreases with the increase of phase adjustment bits. Hence, it is not beneficial to build high-precision elements in practice.}
    \item The upper bound derived from Jensen's inequality and the approximation derived from the channel hardening effect are both proved to be asymptotically equivalent. Moreover, the average achievable rate with correlated channels is asymptotically equivalent to that with uncorrelated channels for a large number of elements. 
    { \item We extend the analysis and study the performance of IOS assisted multi-user NOMA and IOS assisted OMA. We show that, except the user with the last decoding order, the average achievable rates of all other users are determined by the power allocation scheme of the transmitter in the large transmit signal to noise ratio (SNR) regime in a multi-user NOMA system. We also show that IOS-assisted NOMA may not always have higher sum rate than OMA since IOS can reconfigure the channels of users in different time slots of OMA.}
\end{enumerate}

\begin{figure}[t!]
    \centerline{\includegraphics[scale=0.18]{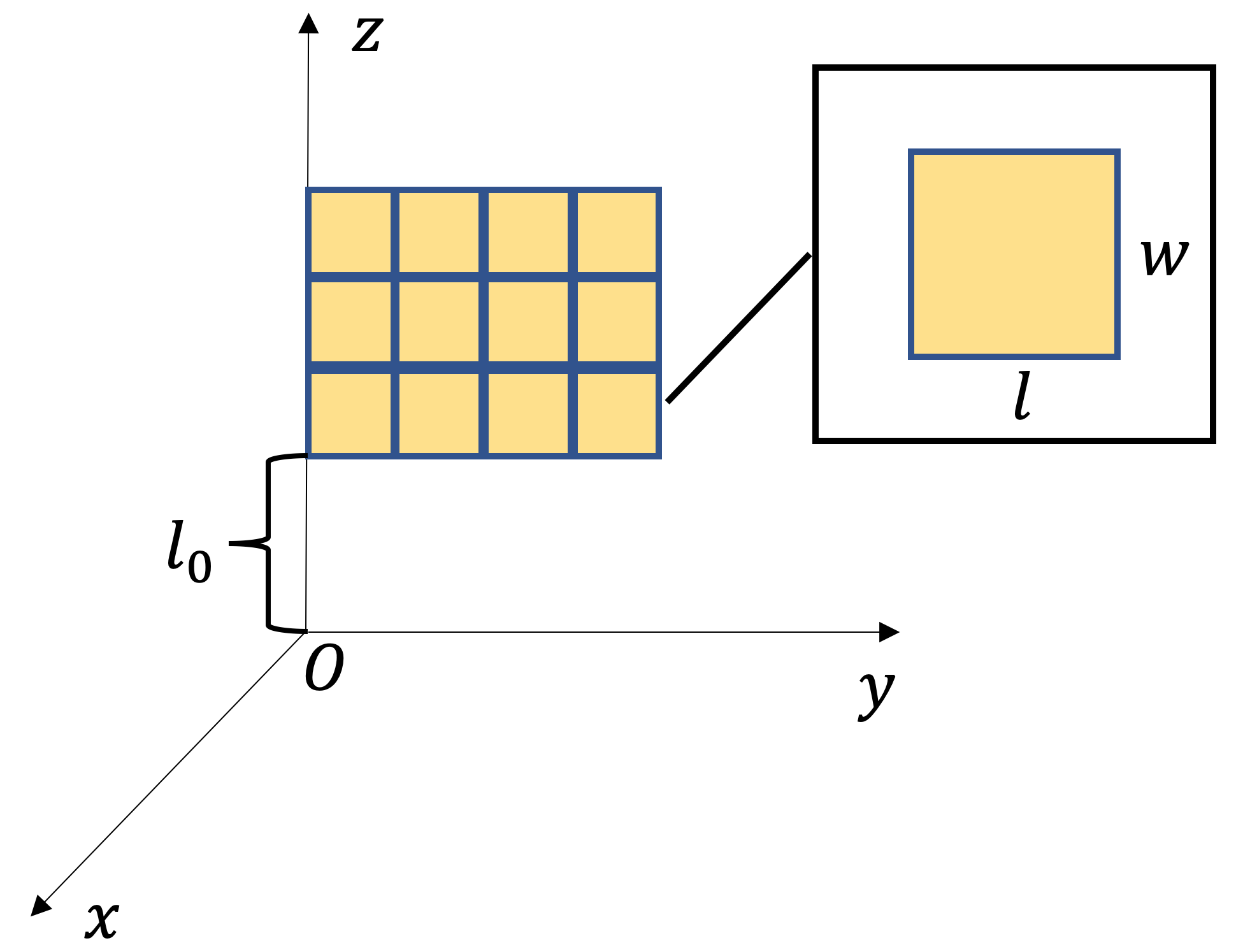}}   
    \caption{IOS structure.}
    \label{sys2}
\end{figure}

\begin{table}[t!]
\renewcommand\arraystretch{1.5}  
    \centering
   \caption{Notations of Rates in Different Scenarios of User $M$. $M \in \{T, R\}$. $m\in\{t, r\}$.}
\begin{tabular}{|c|c|}
    
    \hline
     $R_m$&   Average achievable rate of $M$ under NOMA.\\
     \hline
     $R_m^{*}$&  An Upper bound on $R_m$ by Jensen's inequality.\\
     \hline
     $R_m^{+}$& An approximation to $R_m$ by large array analysis.\\
     \hline
     $R_{m,u}$& $R_m$ with uncorrelated channels.\\
     \hline
     $\Tilde{R}_m$& Average achievable rate of $M$ under OMA.\\
     \hline
     $\Tilde{R}_m^{*}$& An Upper bound on $\Tilde{R}_m$ by Jensen's inequality.\\
     \hline
     $\Tilde{R}_m^{+}$& An approximation to $\Tilde{R}_m$ by large array analysis.\\
     \hline
\end{tabular}
\label{tb11}
\end{table}

This paper is organized as follows. In Section II, we introduce the system model used in this paper. Upper bounds on the average achievable rates of different users are analyzed in Section III. Section IV discusses the channel hardening effect, the asymptotic behavior of the average achievable rates, and the impact of quantization bits. {  IOS assisted multi-user NOMA is analyzed in Section V.} Section VI presents the numerical results, followed by the Conclusion in Section VII. {  For clarity, Table.~\ref{tb11} lists some key notations of rates in different scenarios.}

\textit{Notations}: $\mathbb{E}[X]$ and ${\rm Var}[X]$ denote the expectation and variance of a random variable $X$. ${\rm Cov}\{X, Y\}$ denotes the covariance of random variables $X$ and $Y$. Bold lower-case and bold upper-case letters represent vectors and matrices, respectively. ${\rm mod}(a, b)$ denotes the remainder of division $a$ by $b$. $\lfloor a \rfloor$ represents the integer closet to $a$ that is less than or equal to $a$. $\mathbb{C}^{N \times 1}$ denotes the $N \times 1$ complex vectors space. $\mathbf{A}^{T}$ and $\mathbf{A}^{H}$ respectively stand for the transpose and Hermitian transpose of 
matrix $\mathbf{A}$. ${\rm diag(\mathbf{a})}$ represents a diagonal matrix with the elements of $\mathbf{a}$ on its main diagonal. `$\to$', `$\overset{P}{\to}$' and `$\triangleq$' stand for `approaching to a particular value', `convergence in probability' and `definition', respectively.

\section{System Model}
In this paper, we consider an IOS-assisted communication network, where a single-antenna transmitter (TX) communicates with two single-antenna users. Two users are located on different sides of the IOS. As shown in Fig.~\ref{sys1}, the reflecting user $R$ is located on the same side of the IOS as TX, while the transmitting user $T$ is on the other side. It is assumed that obstacles block the direct links from the TX to the users. This set-up reflects scenarios where IOSs are deployed on the windows or building facades, serving users located inside and outside of the buildings simultaneously.


\subsection{IOS Implementations}
We consider an IOS with $N = N_h N_v$ elements placed in a planar rectangular array. As illustrated in Fig.~\ref{sys2}, the IOS is on the $yOz$ plane with $N_h$ elements per row and $N_v$ elements per column. Without loss of generality, we assume each element is of size $S = l \times w$, where $l$ and $w$ are the horizontal and vertical lengths of each element. The IOS is equipped on a building of height $l_0$, and the coordinate of the $n$th element can be written as
\begin{equation}
    \mathbf{a}_n = \left[0, y(n) l, z(n) w + l_0\right],
\end{equation}
where 
\begin{equation}
    y(n) = {\rm mod}(n-1, N_h),~~~ z(n) = \lfloor \frac{n-1}{N_h} \rfloor.\label{Za}
\end{equation}
$y(n)$ and $z(n)$ are the indexes of the $n$th element on the $y$-axis and $z$-axis.

Each IOS element is able to adjust phase shifts for the user $T$ and user $R$ independently, as introduced in \cite{xu2021star,9570143,wu2021coverage}. Denoting the phase shifts imposed on the incident signals by the $n$th element in the transmitting and reflecting modes as $\theta_n$ and $\psi_n$, the transmitting and reflecting coefficients matrices can be written as

\begin{subequations}
    \begin{align}
        \mathbf{\Theta} &= \alpha\,{\rm diag}\left(e^{j \theta_{1}},..., e^{j \theta_{N}}\right), \\
        \mathbf{\Psi} &= \beta\,{\rm diag}\left(e^{j \psi_{1}},..., e^{j \psi_{N}}\right),
    \end{align}
\end{subequations}
where $\alpha$ and $\beta$ are the transmitting and reflecting amplitude coefficients of the IOS. The IOS is assumed to be passive with no energy dissipation. Hence, the amplitude coefficients of each element satisfy \cite{zhang2022star}
\begin{equation}
    \alpha^2 + \beta^2 = 1.
\end{equation}

\subsection{Channel and Signal Models}
\subsubsection{Small Scale Fading} Let $\mathbf{h} = \left(h_1, ... ,h_N\right)^{T} \in \mathbb{C}^{N\times 1}$, $\mathbf{g} = \left(g_{1}, ... ,g_{N}\right)^{T} \in \mathbb{C}^{N\times 1}$ and $\mathbf{r} = \left(r_{1}, ... ,r_{N}\right)^{T} \in \mathbb{C}^{N\times 1}$ denote the normalized small scale fading of the TX-IOS, IOS-user $T$ and IOS-user $R$ channels, respectively. IOS operates in the far-field of TX and users, thus the EM waves impinging on the IOS are plane waves with the wave vector \cite{bjornson2017massive}
\begin{equation}
    \mathbf{v}(\varrho, \varphi) = \frac{2 \pi}{\lambda}\left[\cos(\varrho) \cos(\varphi), \sin(\varrho)\cos(\varphi), \sin(\varphi)\right],
\end{equation}
where $\varrho$ and $\varphi$ are the azimuth and elevation angles; $\lambda$ is the wavelength. 

Due to the multi-path propagation, the small scale fading $\mathbf{h}$ is a superposition of the independent array response of each multi-path component $\mathbf{c}_k$ as \cite{tse2005fundamentals}
\begin{equation}
    \mathbf{h} = \sum\limits_{k=1}^{K} \mathbf{c}_k  = \sum\limits_{k=1}^{K} \frac{l_k}{\sqrt{K}} \hat{\mathbf{c}}_k,
\end{equation}
where $l_k/\sqrt{K}$ is a complex random variable that stands for the attenuation and phase-rotation of the $k$th multi-path component and $\hat{\mathbf{c}}_k = \left[\exp\left({j \mathbf{v} \mathbf{a}_{1}^{T}}\right), \exp\left({j \mathbf{v} \mathbf{a}_{2}^{T}}\right), ... , \exp\left({j \mathbf{v} \mathbf{a}_{N}^{T}}\right)\right]^{T}$. In non-LoS scenarios, $l_k$, $k \in \{1, 2, ..., K\}$, are independent and identical (i.i.d.) random variables with zero mean and unit variance.
In a rich scattering environment, there are infinite number of multi-path components ($K \to \infty$). Hence, with the help of central limit theorem \cite{bjornson2020rayleigh}, we have\footnote{  This model is valid for rich scattering urban environments without a stable LoS path.}
\begin{equation}
    \mathbf{h} \sim \mathcal{CN}\left(\mathbf{0}_{N}, \mathbf{R}\right),
\end{equation}
where $\mathbf{R}$ is the covariance matrix of $\mathbf{h}$, defined as
\begin{equation}
    \mathbf{R} = \mathbb{E}\left[\mathbf{h} \mathbf{h}^{H}\right] = \mathbb{E}\left[\sum\limits_{k=1}^{K} \mathbf{c}_{k} \mathbf{c}_{k}^{H}\right] = \frac{1}{K} \mathbb{E}\left[\sum\limits_{k=1}^{K} \hat{\mathbf{c}}_{k} \hat{\mathbf{c}}_{k}^{H}\right].
\end{equation}
Using the result in \cite[Prop. 1]{bjornson2020rayleigh}, each element in $\mathbf{R}$ can be written as
\begin{equation}
    \left[\mathbf{R}\right]_{m, n} = \frac{\sin\left(\frac{2 \pi}{\lambda}\parallel \mathbf{a}_m - \mathbf{a}_n \parallel\right)}{\frac{2 \pi}{\lambda}\parallel \mathbf{a}_m - \mathbf{a}_n \parallel}.
    \label{corr_m_n}
\end{equation}
In similar ways, we can have
\begin{subequations}
    \begin{align}
        \mathbf{g} &\sim \mathcal{CN}\left(\mathbf{0}_{N}, \mathbf{R}\right), \\
        \mathbf{r} &\sim \mathcal{CN}\left(\mathbf{0}_{N}, \mathbf{R}\right).
    \end{align}
\end{subequations}

\subsubsection{Large Scale Fading} The pathloss of the TX-IOS-user $T$ and TX-IOS-user $R$ links can be written as \cite{ozdogan2019intelligent}
\begin{equation} 
    \eta_t = \frac{\Lambda_t}{d_b^\chi d_t^\chi}, \, \eta_r = \frac{\Lambda_r}{d_b^\chi d_r^\chi},
\end{equation}
where $\Lambda_t$ and $\Lambda_r$ stand for the intercepts at a reference distance of $1 {\rm m}$ of the TX-IOS-user $T$ and TX-IOS-user $R$ links, respectively; $\chi$ is the pathloss exponent; $d_{b}$, $d_{t}$ and $d_{r}$ denote the distances of the TX-IOS, IOS-user $T$ and IOS-user $R$ links, respectively.

\subsubsection{Signal Models} In the NOMA transmission scheme, TX broadcasts the superimposed signals of the two users\footnote{Note that more than two users can be selected to perform NOMA, but the performance might be degraded since the co-channel interference of the NOMA scheme can be severe and the SIC can be very complex. To optimize the performance of NOMA, hybrid NOMA can be implemented, by combining NOMA with conventional OMA, such as TDMA, FDMA.} as
\begin{equation}
    s = \sqrt{P}\left(q_t s_t + q_r s_r\right),
\end{equation}
where $s_t$ and $s_r$ denote the signals of user $T$ and user $R$ with unit power, i.e., ${\mathbb{E}}[\left|s_t\right|^2] = {\mathbb{E}}[\left|s_r\right|^2] = 1$. $P$ is the total transmit power of TX. $q_t$ and $q_r$ are the transmit amplitude coefficients for user $T$ and user $R$ at TX, satisfying $q_t^2 + q_r^2 = 1$. Hence, the received signals at user $T$ and user $R$ can be written as
\begin{subequations}
    \begin{align}
        y_t &= \sqrt{\eta_t} \mathbf{g}^{T} \mathbf{\Theta} \mathbf{h} \sqrt{P} \left(q_t s_t + q_r s_r\right) + n_t, \label{r_y_t}\\
        y_r &= \sqrt{\eta_r} \mathbf{r}^{T} \mathbf{\Psi} \mathbf{h} \sqrt{P} \left(q_t s_t + q_r s_r\right) + n_r, \label{r_y_r}
    \end{align}
\end{subequations}
where $n_t$ and $n_r$ are the additive white Gaussian noise at user $T$ and user $R$ with zero mean and variance $\sigma_0^2$. 

In the NOMA transmission scheme, one of the two users performs successive interference cancellation (SIC) by decoding the other user's signal, subtracting it from the received signal, and decoding its own signal. The optimal decoding order corresponds to the ascending order of the channel gains of the two users. However, the order of the channel gains is 
dictated by the small scale fading which fluctuates quickly and renders the analysis complicated. It is worth highlighting that we do not aim to 
investigate the optimal decoding order in this work. Alternatively, we assume the distance of the IOS-user $T$ channel is smaller than that of the IOS-user $R$ channel\footnote{  In practice, the locations of different users can be readily obtained from a location information system, such as global positioning system (GPS) \cite{hu2020location}.}, and the decoding order is fixed as ($R$, $T$). Specifically, user $R$ decodes its signal directly by treating the signal of user $T$ as interference, and user $T$ performs SIC. A typical application scenario of IOS is to deploy IOS on the facades of buildings to serve indoor user ($T$) and outdoor user ($R$) simultaneously. Since the indoor user ($T$) is closer to the IOS than the outdoor user ($R$) in general, the large scale fading of the indoor user ($T$) is less severe than that of the outdoor user ($R$), which supports the practicability of the assumption. Furthermore, this assumption is also adopted in many other similar works, such as \cite{zhu2020power,ding2020simple, chen2016application, fang2020energy}. Based on the principles of NOMA transmission scheme, user $R$ is allocated with more transmit power to ensure that user $T$ could perform SIC successfully, i.e., $q_t < q_r$. Hence, signal to interference plus noise ratio (SINR) of user $T$ decoding the signal of user $R$ in the SIC process can be written as
\begin{equation}
    \gamma_{t \to r} = \frac{\gamma_{t \to r}^{(s)}}{\gamma_{t \to r}^{(i)}} = \frac{P q_r^2 \eta_t \left|\mathbf{g}^{T} \mathbf{\Theta} \mathbf{h}\right|^2}{P q_t^2 \eta_t \left|\mathbf{g}^{T} \mathbf{\Theta} \mathbf{h}\right|^2 + \sigma_0^2}.
    \label{snr_t_r}
\end{equation}
After subtracting the signal of user $R$ from the received signal, user $T$ is able to decode its own signal with SNR
\begin{equation}
    \gamma_{t} = \gamma_0 q_t^2 \eta_t \left|\mathbf{g}^{T} \mathbf{\Theta} \mathbf{h}\right|^2,
    \label{snr_t_0}
\end{equation}
where $\gamma_0 = P / \sigma_0^2$ is the transmit SNR at TX. 

On the other hand, user $R$ decodes its own signal directly with SINR
\begin{equation}
    \gamma_{r} = \frac{\gamma_{r}^{(s)}}{\gamma_{r}^{(i)}} = \frac{P q_r^2 \eta_r \left|\mathbf{r}^{T} \mathbf{\Psi} \mathbf{h}\right|^2}{P q_t^2 \eta_r \left|\mathbf{r}^{T} \mathbf{\Psi} \mathbf{h}\right|^2 + \sigma_0^2}.
    \label{snr_r_000}
\end{equation}

\subsubsection{Phase Adjustments}
It can be proved that $\gamma_{t \to r}$, $\gamma_t$ and $\gamma_r$ in \eqref{snr_t_r}, \eqref{snr_t_0} and \eqref{snr_r_000} are maximized when the phase shifts $\theta_n$ and $\psi_n$ are controlled to co-phase the transmitting and reflecting links, respectively \cite{wu2021coverage}. Thus, the optimal phase shifts can be written as
\begin{subequations}
    \begin{align}
        \Bar{\theta}_{n} &= -\arg\left(h_n\right) - \arg\left(g_n\right), \, n \in \{1, ..., N\},\\
        \Bar{\psi}_{n} &= -\arg\left(r_n\right) - \arg\left(g_n\right), \, n \in \{1, ..., N\}.
    \end{align}
\end{subequations}
However, the phase adjustment may not be perfect in practice due to imperfect channel knowledge and phase quantization, which leads to the phase error $\{\phi_{n}^{u}\}_{n=1}^{N}$, $u \in \{t, r\}$ at each IOS element \cite{b8}. These two types of errors are introduced as follows:
\begin{itemize}
    \item Channel estimation errors: Channel estimation may not be precise in practice. And the phase error $\{\phi_{n}^{u}\}_{n=1}^{N}$, $u \in \{t, r\}$ due to inaccurate channel estimation is modeled by i.i.d. Von Mises random variables with zero mean and concentration parameter $\kappa_u$ \cite{b8,wang2021outage}, of which the probability density function (PDF) can be written as
    \begin{equation}
        f_{\phi_n^{u}}(\theta) = \frac{e^{\kappa_u \cos(\theta)}}{2 \pi I_0(\kappa_u)}, \qquad -\pi < \theta < \pi,
    \end{equation}
    where $I_0(\cdot)$ is the modified Bessel function of the first kind of order zero. PDF of $\phi_n^{u}$ is symmetric around zero and gets more concentrated with the increase of $\kappa_u$, which also means the channel estimation is more accurate.
    \item Phase quantization errors: Assuming that each element of the IOS is a $b$-bit phase shifter, the phase error can be modeled by i.i.d. random variables uniformly distributed over $\{-\frac{\pi}{2^b}, \frac{\pi}{2^b}\}$, whose distribution is symmetric around zero \cite{9322575}.
\end{itemize}

Due to phase errors, the realistic phase shifts at each element can be written as
\begin{equation}
    \hat{\theta}_n = \Bar{\theta}_n + \phi_n^{t}, \qquad \hat{\psi}_n = \Bar{\psi}_n + \phi_n^{r}. 
\end{equation}

\subsubsection{Average Achievable Rate} After phase adjustment, $\gamma_{t \to r}$, $\gamma_t$ and $\gamma_r$ can be rewritten as
\begin{subequations}
    \begin{align}
        \gamma_{t} &= \gamma_0 q_t^2 \eta_t \alpha^2 H_t,
        \label{snr_t_0_p} \\[5pt]
        \gamma_{t \to r} &= \frac{\gamma_{t \to r}^{(s)}}{\gamma_{t \to r}^{(i)}} = \frac{P q_r^2 \eta_t \alpha^2 H_t}{P q_t^2 \eta_t \alpha^2 H_t + \sigma_0^2}, \label{snr_t_r_p} \\[5pt]
        \gamma_{r} &= \frac{\gamma_{r}^{(s)}}{\gamma_{r}^{(i)}} = \frac{P q_r^2 \eta_r \beta^2 H_r}{P q_t^2 \eta_r \beta^2 H_r + \sigma_0^2},
        \label{snr_r_0}
    \end{align}
\end{subequations}
where
\begin{equation}
    H_t = \left|\sum\limits_{n=1}^{N}\left|g_{n}\right| \left|h_n\right| e^{j\phi_{n}^{t}}\right|^2,\quad H_r = \left|\sum\limits_{n=1}^{N}\left|r_{n}\right| \left|h_n\right| e^{j\phi_{n}^{r}}\right|^2.
    \label{hthr}
\end{equation}

The average achievable rates of user $T$ and user $R$ can be written as \cite{zhu2020power}
\begin{align}
    R_t &= \mathbb{E}[\log_2(1 + \gamma_t)], \label{R_t_ini}\\
    R_r &= \mathbb{E}\left[\min\left\{\log_2(1 + \gamma_{t \to r}), \log_2(1 + \gamma_{r})\right\}\right]. \label{R_r_in}
\end{align}

{  \subsection{IOS-Assisted OMA}
For the purpose of comparison, the OMA scheme using time division multiple access (TDMA) is introduced here. In the OMA scheme, TX communicates to user $T$ and user $R$ consecutively over two identical time slots. In each time slot, the phase shifts and amplitude coefficients of the IOS are adjusted to enhance the link of one of the users. Hence, the average achievable rate of user $T$ in the OMA scenario are
\begin{equation}
    \Tilde{R}_t = \frac{1}{2} \mathbb{E}[\log_2(1 + \Tilde{\gamma}_t)],
\end{equation}
where $\Tilde{\gamma}_t = \gamma_0 \eta_t H_t$.
Similarly, the average achievable rate of use $R$ is
\begin{equation}
    \Tilde{R}_r = \frac{1}{2} \mathbb{E}[\log_2(1 + \Tilde{\gamma}_r)],
\end{equation}
where $\Tilde{\gamma}_r = \gamma_0 \eta_r H_r$.
}

\section{Average Achievable Rate Analysis}
In this section, we analyze the average achievable rates of user $T$ and user $R$, respectively. 

\subsection{Average Achievable Rate of User $T$}
It is mathematically intractable to derive the exact close-form expression of $R_t$. Alternatively, we use Jensen's inequality to obtain 
an upper bound $R_t^{*}$ as
\begin{equation}
    \begin{split}
        R_t^{*} &= \log_2\left(1 + \mathbb{E}[\gamma_t]\right) \\[5pt]
        &= \log_2\left(1 + \gamma_0 q_t^2 \eta_t \alpha^2 \mathbb{E}\left[H_t\right] \right),
    \end{split}
    \label{approx_r_t_0}
\end{equation}
where $\mathbb{E}\left[H_t\right]$ can be further decomposed as \eqref{decom_t} shown at the top of the next page.
\begin{figure*}
    \begin{equation}
        \begin{split}
            \mathbb{E}\left[H_t\right] = &\mathbb{E}\left[\sum\limits_{n=1}^{N} \left|g_{n}\right|^2 \left|h_n\right|^2 \right] + 2\, \mathbb{E}\left[\sum\limits_{n=1}^{N-1}\sum\limits_{i=n+1}^{N} \left|g_{n}\right|\left|h_n\right| \cos(\phi_{n}^{t}) \left|g_{i}\right|\left|h_i\right| \cos(\phi_{i}^{t})\right] \\
            &+ 2\, \mathbb{E}\left[\sum\limits_{n=1}^{N-1}\sum\limits_{i=n+1}^{N} \left|g_{n}\right|\left|h_n\right| \sin(\phi_{n}^{t}) \left|g_{i}\right|\left|h_i\right| \sin(\phi_{i}^{t})\right]
        \end{split}
        \label{decom_t}
    \end{equation}
    \hrule
\end{figure*}
Since the channels $\mathbf{h}$ and $\mathbf{g}$ are correlated, we need to figure out $\mathbb{E}\left[\left|h_n\right|\left|h_i\right|\right]$ and $\mathbb{E}\left[\left|g_n\right|\left|g_i\right|\right]$. The result is summarized in the following proposition.
\begin{proposition}
    For $n \neq i$, $\mathbb{E}\left[\left|w_n\right|\left|w_i\right|\right]$, $w \in \{h, g, r\}$ can be expressed in terms of $\mathbb{E}\left[w_n w_i^{*}\right]$ as
    \begin{align}
        & \mathbb{E}\left[\left|w_n\right|\left|w_i\right|\right] = \nonumber\\
        & \left(\frac{\left|\mathbb{E}[w_n w_i^{*}]\right|^2}{2} - \frac{1}{2}\right) K\left({\left|\mathbb{E}[w_n w_i^{*}]\right|^2}\right) +  E\left({\left|\mathbb{E}[w_n w_i^{*}]\right|^2}\right). \label{exp_e_h_h}
    \end{align}
    For $n = i$, 
    \begin{equation}
        \mathbb{E}\left[\left|w_n\right|\left|w_i\right|\right] = 1.
    \end{equation}
    $K(\cdot)$ and $E(\cdot)$ are the complete elliptic integral of the first kind and the second kind, respectively.
    \label{prop11}
\end{proposition}
\begin{IEEEproof}
    See Appendix A.
\end{IEEEproof}
\begin{corollary}
    Under the proposed system settings, $\frac{\pi}{4} \leq \mathbb{E}\left[\left|w_n\right|\left|w_i\right|\right] \leq 1$, for all $n,i$. 
    \label{corr11}
\end{corollary}
\begin{IEEEproof}
    See Appendix B.
\end{IEEEproof}

Since the covariance matrices of $\mathbf{h}$ and $\mathbf{g}$ are equal, it can be learned that the matrices $\mathbb{E}[|\mathbf{h}||\mathbf{h}|^{T}]$ and $\mathbb{E}[|\mathbf{g}||\mathbf{g}|^{T}]$ are equal, and we denote each by $\Bar{\mathbf{R}}$. Applying Proposition \ref{prop11}, 
an upper bound on the average achievable rate of user $T$ can be obtained, which is summarized in the following proposition.
\begin{proposition}
    An upper bound on the average achievable rate of user $T$ can be written as
    \begin{equation}
        R_t^{*} = \log_2\left(1 + \gamma_0 q_t^2 \eta_t \alpha^2 \left(N (1 - \epsilon_t^2) + \epsilon_t^2 \mathbf{tr}\left(\mathbf{\Bar{R} \Bar{R}}\right)\right) \right),
        \label{r_t_*}
    \end{equation}
    where
    \begin{equation}
        \epsilon_t = \mathbb{E}{\left[\cos(\phi_{n}^{t})\right]},\, n \in\{ 1, 2, ..., N\}.
    \end{equation}
    For phase estimation errors, 
    \begin{equation}
        \epsilon_t = \frac{I_1(\kappa_t)}{I_0(\kappa_t)}.
        \label{etvm}
    \end{equation}
    For phase quantization errors,
    \begin{equation}
        \epsilon_t = \frac{2^b \sin{\left(\frac{\pi}{2^b}\right)}}{\pi}.
    \end{equation}
    \label{prop22}
\end{proposition}
\begin{IEEEproof}
    See Appendix C.
\end{IEEEproof}

\begin{remark}
    For perfect phase adjustment, i.e., $\phi_{n}^{t} = 0$, $n \in\{ 1, 2, ..., N\}$, the 
    upper bound on the average achievable rate of user $T$ is
    \begin{equation}
        R_t^{*} = \log_2\left(1 + \gamma_0 q_t^2 \eta_t \alpha^2 \mathbf{tr}\left(\mathbf{\Bar{R} \Bar{R}}\right) \right).
    \end{equation}
\end{remark}
\begin{remark}
    When the phase errors are uniformly distributed over $[-\pi, \pi)$, i.e., $\epsilon_t = 0$ , the 
    upper bound on the average achievable rate of user $T$ is
    \begin{equation}
        R_t^{*} = \log_2\left(1 + \gamma_0 q_t^2 \eta_t \alpha^2 N \right).
    \end{equation}
    It can be seen that $R_{t}^{*}$ scales with $\log_2(N)$ when the phase errors are uniformly distributed over $[-\pi, \pi)$, which coincides with the proposition in \cite{8811733}.
\end{remark}

\begin{corollary}
    Under the proposed system settings, $\mathbb{E}[H_1]$ 
    satisfies
    \begin{equation}
        N + \frac{\pi^2 N (N-1)}{16} \epsilon_t^2\,{\leq}\,\mathbb{E}[H_1]\,{<}\,N + N (N-1) \epsilon_t^2.
    \end{equation}
\end{corollary}
\begin{IEEEproof}
    The conclusion is obtained by substituting the result in Corollary \ref{corr11} into \eqref{decom_t}.
\end{IEEEproof}

\begin{remark}
    It can be learned from Corollary 2 that $R_t^{*}$ scales with $\log_2(N^2)$ when $\epsilon_t \neq 0$. Compared with Remark 2, it can be seen that even imperfect phase adjustment is able to provide performance gain, which scales with $\log_2(N)$. This is a known result for the system with uncorrelated channels \cite{b8}, and it is shown here that it still works for correlated channels. Simulation results will illustrate this point in Section V.
    \label{remm33}
\end{remark}

{  \begin{corollary}
    Following the similar procedures in the Propositions \ref{prop11} and \ref{prop22}, upper bounds on the average achievable rates of user $T$ and user $R$ by using OMA scheme is given by
    \begin{subequations}
        \begin{align}
            \Tilde{R}_{t}^{*} &= \frac{1}{2} \log_2\left(1 + \gamma_0 \eta_t \left(N (1 - \epsilon_t^2) + \epsilon_t^2 \mathbf{tr}\left(\mathbf{\Bar{R} \Bar{R}}\right)\right) \right),\\
            \Tilde{R}_{r}^{*} &= \frac{1}{2} \log_2\left(1 + \gamma_0 \eta_r \left(N (1 - \epsilon_r^2) + \epsilon_r^2 \mathbf{tr}\left(\mathbf{\Bar{R} \Bar{R}}\right)\right) \right).
        \end{align}
    \end{subequations}
\end{corollary}}

\subsection{Average Achievable Rate of User $R$}
It appears that obtaining the close-form expression of $R_r$ in \eqref{R_r_in} is intricate. Alternatively, we first analyze $R_r$ for large transmit SNR $\gamma_0$ and then seek for an upper bound based on Jensen's inequality.  

\subsubsection{Analysis of Large Transmit SNR} 
$\gamma_{t \to r}$, defined in \eqref{snr_t_r_p}, can be rewritten as
\begin{equation}
    \gamma_{t \to r} = \frac{q_r^2}{q_t^2 + \left(\gamma_0 \eta_t \alpha^2 H_t\right)^{-1}}.
\end{equation}
As $\gamma_0 \to \infty$, we can have
\begin{equation}
    \gamma_{t\to r} \to \frac{q_r^2}{q_t^2}.
\end{equation}
The same conclusion can be derived for $\gamma_r$: $\gamma_{r} \to {q_r^2}/{q_t^2}$ as $\gamma_0 \to \infty$.
Hence,
\begin{equation}
    R_r \to \log_2\left(1 + \frac{q_r^2}{q_t^2}\right),~~~\gamma_0 \to \infty.
    \label{asy_r_r_snr}
\end{equation}

\subsubsection{Analysis of Finite Transmit SNR}
The $\min$ function is concave \cite{boyd2004convex}. Therefore, Jensen's inequality can be applied on $R_r$ to obtain an upper bound on $R_r$ as
\begin{equation}
    R_r \leq \min\left\{\log_2(1 + \mathbb{E}[\gamma_{t \to r}]), \log_2(1 + \mathbb{E}[\gamma_{r}])\right\}.
    \label{R_r_fb}
\end{equation}
Referring to \eqref{snr_t_r_p} and \eqref{snr_r_0}, \eqref{R_r_fb} is hard to 
deal with due to the common terms in the numerators and denominators of $\gamma_{t \to r}$ and $\gamma_{r}$. Specifically, the common term in the numerator and denominator of $\gamma_{t \to r}$ is $H_t$, and the common term of $\gamma_{r}$ is $H_r$. Fortunately, it can be proved that $\gamma_{t \to r}$ and $\gamma_{r}$ are concave to their respective common terms. Thus, Jensen's inequality can be applied again, which gives
\begin{equation}
    R_r \leq \min\left\{\log_2\left(1 + \frac{\mathbb{E}[\gamma_{t \to r}^{s}]}{\mathbb{E}[\gamma_{t \to r}^{i}]}\right), \log_2\left(1 + \frac{\mathbb{E}[\gamma_{r}^{s}]}{\mathbb{E}[\gamma_{r}^{i}]}\right)\right\}.
\end{equation}
Following the similar procedures in the Propositions \ref{prop11} and \ref{prop22}, an upper bound $R_r^{*}$ can be written as
\begin{equation}
    \begin{split}
        &R_r^{*} = \\
        &\min\left\{\log_2\left(1 + \frac{q_r^2}{q_t^2 + f_t^{-1} }\right), \log_2\left(1 + \frac{q_r^2}{q_t^2 + f_r^{-1} }\right)\right\} \\[5pt]
        &= \left\{
        \begin{split}
            &\quad \log_2\left(1 + \frac{q_r^2}{q_t^2 + f_t^{-1} }\right),\quad f_t < f_r, \\[5pt]
            &\quad \log_2\left(1 + \frac{q_r^2}{q_t^2 + f_r^{-1} }\right),\quad f_t \geq f_r,
        \end{split}
        \right. 
    \end{split}
    \label{r_r_*}
\end{equation}
where
\begin{subequations}
    \begin{align}
        \epsilon_r &= \mathbb{E}{\left[\cos(\phi_{n}^{r})\right]},\, n \in \{1, 2,..., N\},\\[5pt]
        f_t &=  \gamma_0 \eta_t \alpha^2 \left(N (1 - \epsilon_t^2) + \epsilon_t^2 \mathbf{tr}\left(\mathbf{\Bar{R} \Bar{R}}\right)\right), \label{f_t_1}\\[5pt]
        f_r &= \gamma_0 \eta_r \beta^2 \left(N (1 - \epsilon_r^2) + \epsilon_r^2 \mathbf{tr}\left(\mathbf{\Bar{R} \Bar{R}}\right)\right).\label{f_r_1}
    \end{align}
\end{subequations}

\section{Large Array Analysis}
In this section, we 
show that the channel hardening effect appears in the proposed system and revisit the average achievable rates 
for a large number of IOS elements. 

\subsection{Channel Hardening Analysis}
As defined in \cite{bjornson2020rayleigh}, the channel hardening effect in a RIS-assisted network refers that the received SNR variations average out and the received SNR is approximately equal to $N^2$ times a constant as $N$ goes large. The channel hardening effect in the proposed system model with correlated channels and imperfect phase adjustment is presented in the following proposition.
\begin{proposition}
    The composite channel gains associated with user $T$ and user $R$, namely $H_t$ and $H_r$, satisfy
    \begin{subequations}
        \begin{align}
            \frac{1}{N^2} H_t \overset{P}{\to} \frac{\pi^2}{16} \epsilon_t^2,\,\, N \to \infty, \label{h_t_n^2}\\[5pt]
            \frac{1}{N^2} H_r \overset{P}{\to} \frac{\pi^2}{16} \epsilon_r^2,\,\, N \to \infty, \label{h_r_n^2}
        \end{align}
    \end{subequations}
    with the convergence in probability. 
    Hence, for non-uniform phase errors over $[-\pi, \pi)$, $H_t$ and $H_r$ can be approximated by a constant as
    \begin{equation}
        H_t \approx \frac{\pi^2}{16} N^2 \epsilon_t^2,~~~ H_r \approx \frac{\pi^2}{16} N^2 \epsilon_r^2,
        \label{ch_h_t_new}
    \end{equation}
    when the number of IOS elements is large.
    \label{prop33}
\end{proposition}
\begin{IEEEproof}
    See Appendix D.
\end{IEEEproof}
\begin{remark}
    Under the correlation profile proposed in \eqref{corr_m_n}, the channel correlation does not play a role in the conclusion of Proposition \ref{prop33}. And, in the case of uncorrelated channels, the same approximations on $H_t$ and $H_r$ 
    can derived by using the conventional law of large numbers.
    \label{rem4}
\end{remark}


\subsection{Average Achievable Rates for Large Number of Elements}
We revisit the average achievable rates when the number of elements is large. By using \eqref{ch_h_t_new}, 
the average achievable rates of user $T$ and user
$R$ with non-uniform phase errors over $[-\pi,\pi)$ can be 
approximated as
\begin{equation}
    R_t \approx R_t^{+} = \log_2\left(1 +  \frac{\pi^2 N^2 \gamma_0  \epsilon_t^2 q_t^2 \eta_t \alpha^2}{16}\right), \label{r_t_n_large}
\end{equation}
\begin{equation}
    R_r \approx R_r^{+} =
    \left\{
    \begin{split}
        & \log_2\left(1 + \frac{q_r^2}{q_t^2 + 16 (\pi^2 N^2 \gamma_0 \epsilon_t^2 \eta_t \alpha^2)^{-1}}\right), \\[5pt]
        &\qquad\qquad\qquad\qquad\qquad~~ \epsilon_t^2 \eta_t \alpha^2 < \epsilon_r^2 \eta_r \beta^2, \\[5pt]
        & \log_2\left(1 + \frac{q_r^2}{q_t^2 + 16 (\pi^2 N^2 \gamma_0 \epsilon_r^2 \eta_r \beta^2)^{-1}}\right),\\[5pt]
        &\qquad\qquad\qquad\qquad\qquad~~ \epsilon_t^2 \eta_t \alpha^2 \geq \epsilon_r^2 \eta_r \beta^2.
    \end{split}
    \right. 
    \label{r_r_n_large}
\end{equation}

\begin{remark}
    The impact of phase quantization bits on the average achievable rate can be analyzed by using the result in \eqref{r_t_n_large}. The average achievable rate improvement of user $T$ brought by adding one bit for phase quantization can be expressed as
    \begin{equation}
        R_{b} = \log_2\left(\frac{16 + \pi^2 N^2 \gamma_0  \epsilon_{b+1}^2 q_t^2 \eta_t \alpha^2}{16 + \pi^2 N^2 \gamma_0 \epsilon_{b}^2 q_t^2 \eta_t \alpha^2}\right),
    \end{equation}
    where
    \begin{equation}
        \epsilon_b = \frac{2^b \sin{\left(\frac{\pi}{2^b}\right)}}{\pi}.
    \end{equation}
    For large number of elements, we can have
    \begin{equation}
        R_b \sim f(b),~~~N \to \infty,
    \end{equation}
    where
    \begin{equation}
        f(b) = \log_2\left(\frac{4 \sin^2{\left(\frac{\pi}{2^{b+1}}\right)}}{\sin^2{\left(\frac{\pi}{2^b}\right)}}\right).
    \end{equation}
    $f(b)$ is always positive. However,
    \begin{equation}
        f'(b) = -\frac{\pi}{2^b} \tan \left(\frac{\pi}{2^{b+1}}\right) < 0.
    \end{equation}
    Therefore, adding the number of quantization bits can always improve the average achievable rate of user $T$, but the rate of improvement decreases.
    
    {  The impact of inaccurate channel estimation can also be studied. According to \eqref{etvm}, $\epsilon_u$, $u\in\{t, r\}$ is an increasing function to $\kappa_u$. Further, $R_{u}^{+}$ is also an increasing function to $\epsilon_u$. Therefore, we can learn that more accurate channel estimation (larger $\kappa_u$) will lead to larger average achievable rates of both users.}
    \label{remm55}
\end{remark}

{  \begin{corollary}
In the OMA scheme, the average achievable rates of user $T$ and user $R$ are approximated by
\begin{subequations}
    \begin{align}
        \Tilde{R}_t \approx \Tilde{R}_t^{+} = \frac{1}{2}\log_2\left(1 +  \frac{\pi^2 N^2 \gamma_0  \epsilon_t^2 \eta_t}{16}\right),\label{544a}\\
        \Tilde{R}_r \approx \Tilde{R}_r^{+} = \frac{1}{2}\log_2\left(1 +  \frac{\pi^2 N^2 \gamma_0  \epsilon_r^2 \eta_r}{16}\right).\label{544b}
    \end{align}
\end{subequations}
\end{corollary}}

{  Compared \eqref{r_t_n_large} with \eqref{544a}, we can see that $T$ obtains a higher SNR in OMA at the penalty of a $1/2$ pre-log factor. It can be computed that  ${R}_{t}^{+} > \Tilde{R}_{t}^{+}$ if 
\begin{equation}
    \gamma_0 > \frac{16-32 q_t^2 \alpha^2}{\pi^2 N^2 \epsilon_t^2 \eta_t q_t^4 \alpha^4}.
\end{equation}
Compared \eqref{r_r_n_large} with \eqref{544b}, it can be learned that $\Tilde{R}_{r}^{+} > {R}_{r}^{+}$ in the large transmit SNR regime, since ${R}_{r}^{+}$ converges to a constant as $\gamma_0$ goes large.

Due to the reconfigurability of IOS in different time slots of OMA, IOS assisted NOMA may not always achieve better performance than NOMA in terms of the system sum rate. Actually, which technique is preferable depends on the system setup. 
    The approximate sum rates of NOMA and OMA, i.e, $S_{NOMA}$ and $S_{OMA}$, is given by
    \begin{subequations}
        \begin{align}
            S_{NOMA} &= R_t^{+} + R_r^{+},\\
            S_{OMA} &= \Tilde{R}_t^{+} + \Tilde{R}_r^{+}.
        \end{align}
    \end{subequations}
    After some mathematical manipulations, it can be derived that $S_{NOMA}$ and $S_{OMA}$ comply with the following relationship in the large transmit SNR regime:
    \begin{equation}
    \left\{
        \begin{split}
            & S_{NOMA} > S_{OMA},~~~ \alpha^4 \epsilon_t^2 \eta_t > \epsilon_r^2 \eta_r, \\[5pt]
            & S_{NOMA} < S_{OMA},~~~\alpha^4 \epsilon_t^2 \eta_t < \epsilon_r^2 \eta_r.
        \end{split}
        \right. 
    \end{equation}

Contrary to the conventional idea that NOMA has superiority over OMA in the conventional wireless system without IOS, NOMA does not always perform better than OMA in the IOS-assisted system due to the fact that IOS is able to reconfigure the channel in each time slot of OMA. Namely, in the time slot for $T$, the amplitude coefficients of IOS is set to be $\alpha = 1$ and $\beta = 0$, and in the time slot for $R$, $\alpha = 0$ and $\beta = 1$. Our analysis shows that, for large transmit SNR, which transmission scheme is better depends on the pathloss, amplitude coefficients at the IOS, and the phase errors distribution. NOMA performs better when the channel of $T$ is remarkably stronger than the channel of $R$. This corresponds to the conclusion in \cite{zheng2020intelligent}. We will corroborate this analysis with numerical results later.}

\subsection{Asymptotic Analysis}
Relations between $R_t^{*}$ and $R_t^{+}$ as well as between $R_r^{*}$ and $R_r^{+}$ are analyzed and summarized in the following proposition.
\begin{proposition}
    For large number of IOS elements with non-uniform phase errors over $[-\pi,\pi)$, 
    the following asymptotic equivalence relations hold:
    \begin{subequations}
        \begin{align}
            R_{t}^{*} &\sim R_t^{+},~~~ N\to \infty,\label{r_t*4a}\\ 
            R_{r}^{*} &\sim R_r^{+},~~~ N\to \infty.\label{r_t*4b}
        \end{align}
    \end{subequations}
    \label{prop4}
\end{proposition}
\begin{IEEEproof}
    See Appendix E.
\end{IEEEproof}
Next, we consider the relation between $R_t$ and $R_t^{*}$.
\begin{proposition}
    For large number of IOS elements with non-uniform phase errors over $[-\pi,\pi)$, $R_t$ and $R_t^{*}$ satisfy the asymptotic equivalence, i.e., 
    \begin{equation}
        R_t \sim R_t^{*},~~~ N \to \infty.
        \label{r_t_r_t*}
    \end{equation}
    Further, due to the transitive property of asymptotic analysis and Proposition \ref{prop4}, $R_t$ and $R_t^{+}$ are also asymptotically equivalent, i.e.,
    \begin{equation}
        R_t \sim R_t^{+},~~~ N \to \infty.
    \end{equation}
    \label{prop5}
\end{proposition}
\begin{IEEEproof}
    See Appendix F.
\end{IEEEproof}

\begin{remark}
    Eq.\eqref{r_t_r_t*} in Proposition \ref{prop5} does not 
    hold when the phase errors are uniformly distributed over $[-\pi, \pi)$. In this case, $R_t^{*}$ only acts as an upper bound to $R_t$. The reason is clarified in Appendix F.
    \label{rem7}
\end{remark}

\begin{remark}
    Due to changing the order of computing expectation and minimum, the relation between $R_r$ and $R_r^{*}$ cannot be analyzed similarly. However, for large number of elements, $R_r$ can be well approximated by $\log_2\left(1 + q_r^2/q_t^2\right)$.
\end{remark}

With the conclusions in Propositions \ref{prop4} and \ref{prop5}, the relation between the average achievable rates of user $T$ with correlated channels, $R_t$, and uncorrelated channels, $R_{t, u}$, is presented in the following proposition.
\begin{proposition}
    For large number of IOS elements with non-uniform  phase errors over $[-\pi,\pi)$, $R_t$ is asymptotically equivalent to $R_{t, u}$, i.e.,
    \begin{equation}
        R_t \sim R_{t, u},~~~ N \to \infty.
    \end{equation}
    \label{prop66}
\end{proposition}

\begin{IEEEproof}
    Denote the approximate expression of $R_{t, u}$ derived from the channel hardening effect as $R_{t, u}^{+}$. Referring to Remark \ref{rem4}, we have $R_t^{+} = R_{t, u}^{+}$. Then, following the similar procedures in Propositions \ref{prop4} and \ref{prop5}, we can obtain $R_t \sim R_{t, u}$ as $N \to \infty$.  
\end{IEEEproof}

{ \begin{remark}
    Since $\Tilde{R}_t$ and $\Tilde{R}_r$ have the same form as $R_t$, the conclusions in the Propositions \ref{prop4}, \ref{prop5} and \ref{prop66} can also be derived for the OMA scenario.
\end{remark}}

{ \section{Discussion on Multiple Users Scenario}
The above analysis can be extended to the multi-user NOMA scenario, where multiple users are located on both sides of the IOS. In this section, we consider a four-user scenario, with two users located on each side of the IOS, i.e., $\mathcal{U}_t\in\{T, T'\}$ and $\mathcal{U}_r\in\{R, R'\}$, where $\mathcal{U}_t$ and $\mathcal{U}_r$ are the collections of users in the transmitting and reflecting regions, respectively. Similar results can be derived using the same method for more users. Without loss of generality, assuming the path loss coefficients of the four users are ordered as 
\begin{equation}
    \eta_{r'} < \eta_{t'} < \eta_{r} < \eta_{t},
\end{equation}
the decoding order is fixed as $(R', T', R, T)$. The transmit amplitude coefficients at TX are $q_{r'}$, $q_{t'}$, $q_{r}$, $q_{t}$, with $q_{r'}^2 + q_{t'}^2 + q_{r}^2 + q_{t}^2 = 1$. By using the priority oriented NOMA scheme proposed in \cite{tahir2020analysis, hou2020reconfigurable}, the transmitting and reflecting phase shifts are adjusted to boost the channels of $T$ and $R$, respectively. The average achievable rates of the four users are given by
\begin{subequations}
    \begin{align}
        R_{t'} &= \mathbb{E}[\min\{\log_2(1+\gamma_{t'}), \log_2(1+\gamma_{r \to t'}), \nonumber\\
        &~~~~~~~~~~ \log_2(1+\gamma_{t \to t'})\}]\\
        R_{r'} &= \mathbb{E}[\min\{\log_2(1+\gamma_{r'}), \log_2(1+\gamma_{t' \to r'}), \nonumber\\
        &~~~~~~~~~~\log_2(1+\gamma_{r \to r'}), \log_2(1+\gamma_{t \to r'})\}],
    \end{align}
\end{subequations}
and $R_t$ and $R_r$ are given in \eqref{R_t_ini} and \eqref{R_r_in}, respectively. Following the similar procedures in Sec. III, closed-form upper bounds of $R_{r'}$ and $R_{t'}$ is formulated in the following proposition.
\begin{proposition}
    Upper bounds on the average achievable rates of $T'$ and $R'$ are given by
    \begin{equation}
        \begin{split}
            R_{t'}^{*} = \left\{
            \begin{split}
                & \log_2\left(1 + \frac{q_{t'}^2}{q_t^2 + q_r^2 + f_{t'}^{-1} }\right),\quad f_{t'} < f_{r}, \\[5pt]
                &\log_2\left(1 + \frac{q_{t'}^2}{q_t^2 + q_r^2 + f_{r}^{-1} }\right),\quad f_{t'} \geq f_{r},
            \end{split}
            \right. 
        \end{split}
        \label{r_t_4_*}
    \end{equation}
    \begin{equation}
        R_{r'}^{*} = \left\{
        \begin{split}
            & \log_2\left(1 + \frac{q_{r'}^2}{q_t^2 + q_r^2 + q_{t'}^2 + f_{t'}^{-1} }\right),\quad f_{t'} < f_{r'}, \\[5pt]
            &\log_2\left(1 + \frac{q_{r'}^2}{q_t^2 + q_r^2 + q_{t'}^2 + f_{r'}^{-1} }\right),\quad f_{t'} \geq f_{r'},
        \end{split}
        \right. 
        \label{r_r_4_*}
    \end{equation}
    where $f_{t}$ and $f_{r}$ is defined in \eqref{f_t_1}and \eqref{f_r_1}; $f_{t'} = \gamma_0 \eta_{t'} \alpha^2 N$ and $f_{r'} = \gamma_0 \eta_{r'} \beta^2 N$.
    \end{proposition}
    \begin{IEEEproof}
        See Appendix G.
    \end{IEEEproof}
    \begin{remark}
        Referring to Appendix G, it can be learned that 
        \begin{subequations}
            \begin{align}
                R_{t'} &\to \log_2\left(1 + \frac{q_{t'}^2}{q_t^2 + q_r^2}\right),~~~ \gamma_0 \to \infty, \label{asy_r_t'_snr}\\
                R_{r'} &\to \log_2\left(1 + \frac{q_{r'}^2}{q_t^2 + q_r^2 + q_{t'}^2}\right),~~~ \gamma_0 \to \infty. \label{asy_r_r'_snr}
            \end{align}
        \end{subequations}
        \label{remm99}
    \end{remark}
    \begin{remark}
        With \eqref{r_t_*}, \eqref{asy_r_r_snr}, \eqref{asy_r_t'_snr} and \eqref{asy_r_r'_snr}, we can conclude that, in NOMA transmission scheme, except the user of the last order (which is $T$ in this paper), the average achievable rates of all other users are determined by the transmit amplitude coefficients of TX as $\gamma_0$ goes large."
    \end{remark}

}

\section{Numerical Results}
In this section, we present numerical results to evaluate the accuracy of the proposed analysis by Monte Carlo (MC) simulations and show the impacts of phase errors and channel correlations on the system performance.

\subsection{Simulation Setup}
Unless otherwise stated, the simulation parameters are set as follows. The distances of the TX-IOS, IOS-user $T$ and IOS-user $R$ links are set to be $d_b = 10\,{\rm m}$, $d_t = 5\,{\rm m}$ and $d_r = 10\,{\rm m}$. The pathloss exponent $\chi$ is $2.4$ and the carrier frequency is $3\,{\rm GHz}$ with the wavelength $\lambda = 0.1\,{\rm m}$. Each IOS element is of size $l = \frac{\lambda}{2}$ and $w = \frac{\lambda}{2}$. {  The transmitting and reflecting amplitude coefficients of the IOS are given as $\alpha = 0.8$ and $\beta = 0.6$, such that $\alpha^2 + \beta^2 = 1$. The transmit amplitude coefficients at TX are $q_t = 0.6$ and $q_r = 0.8$, such that $q_t^2 + q_r^2 = 1$.}\footnote{  In practice, $\alpha$, $\beta$, $q_t$ and $q_r$ can be adjusted to meet different performance requirement of each user. For example, if $R$ needs better performance, we can increase $q_r$ and $\beta$.} The intercepts at unit distance of the transmitting and reflecting channels are $\Lambda_t = \Lambda_r = -30\,{\rm dB}$. The noise power is $\sigma_0^2 = -50\,{\rm dBm}$. 

\subsection{Average Achievable Rates versus Number of IOS Elements}
\begin{figure}[!]
    \centerline{\includegraphics[scale=0.62]{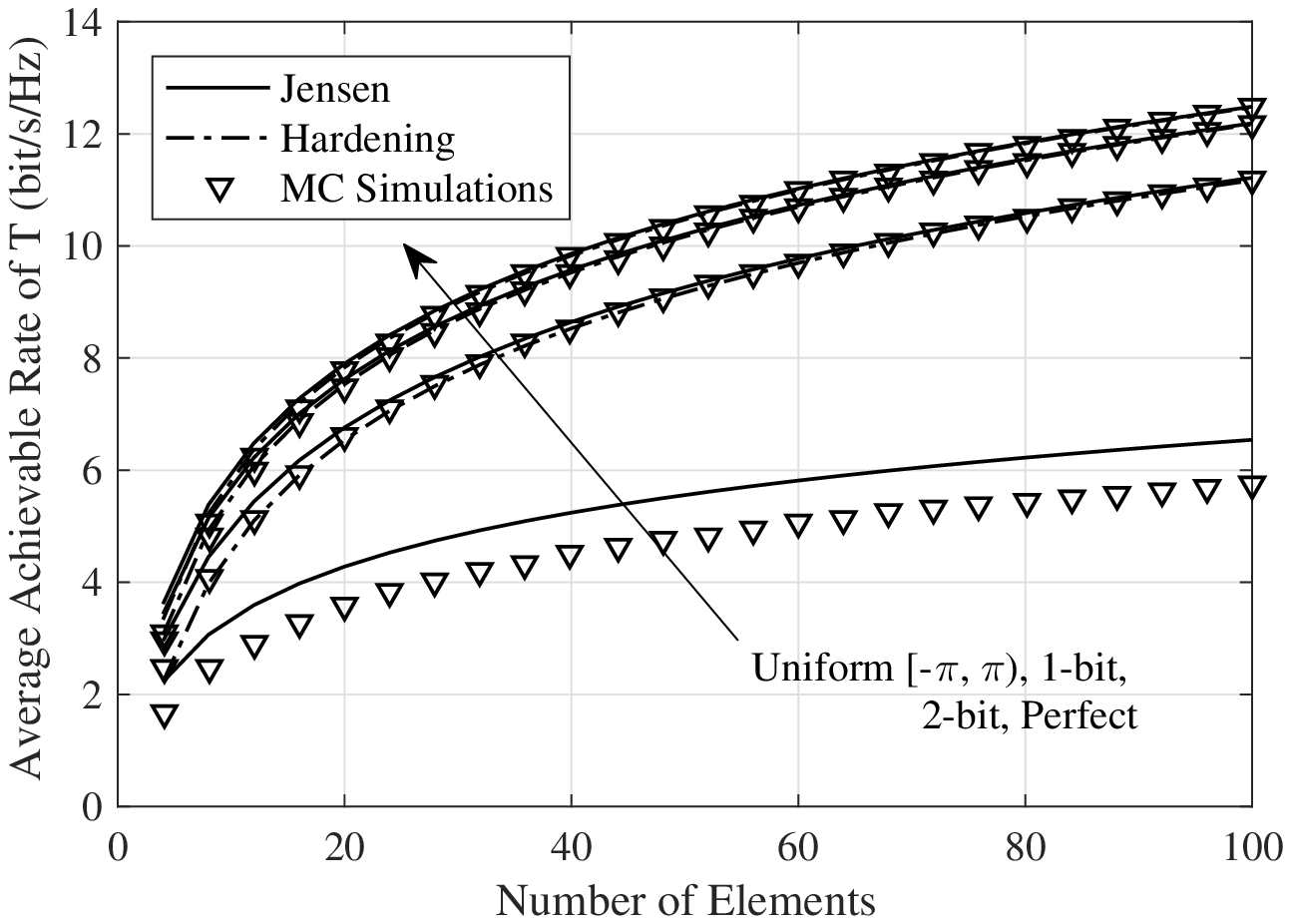}}   
    \caption{Average achievable rate of user $T$ versus number of elements.}
    \label{f1_v2}
\end{figure}
\begin{figure}[!]
    \centerline{\includegraphics[scale=0.58]{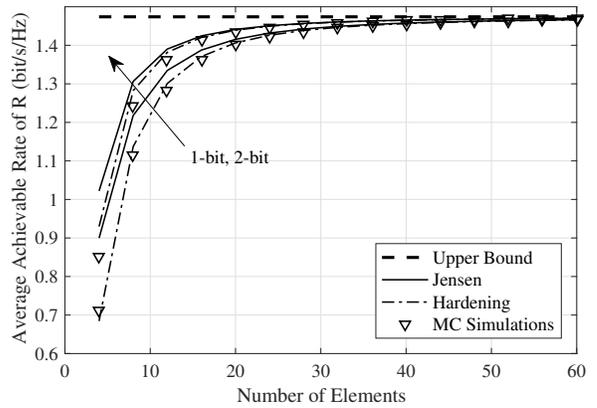}}   
    \caption{Average achievable rate of user $R$ versus number of elements.}
    \label{f2_v2}
\end{figure}
We investigate the average achievable rates versus the number of IOS elements with different phase quantization bits. IOS elements are assumed to be placed in a uniform rectangular array with four elements per column, i.e., $N_v = 4$. The number of elements per row varies in $N_h \in [1, 25]$. The transmit power is set to be $20\, {\rm dBm}$. As benchmarks, the systems with perfect phase adjustment, i.e., $\phi_{n}^{u} = 0$, and with uniform phase error over $[-\pi, \pi)$, are both shown. In the legend, `Jensen' and `Hardening' represent the upper bound and approximation derived in Sec. III and Sec. IV; `MC Simulations' denotes the MC simulation results.

As shown in Fig.~\ref{f1_v2}, the proposed upper bound and approximation on $R_t$ are very accurate even for a small number of elements. Increasing the number of elements can increase the average achievable rate of user $T$, but the slope of the curve becomes flat, which can be concluded from the log relation between $R_t$ and $N^2$ in Remark \ref{remm33}. As expected in Remark \ref{remm55}, adding the number of quantization bits is able to improve system performance. However, the performance gain diminishes with the increase of quantization levels. For example, when the number of elements is 60, the average achievable rates of user $T$ are 5 bit/s/Hz and 9.7 bit/s/Hz for uniform phase errors and 1-bit phase shifting, respectively. However, compared with 1-bit phase shifting, the performance gain brought by 2-bit phase shifting is much smaller with 10.7 bit/s/Hz, and it is very close to the performance of perfect phase shifting with 11 bit/s/Hz. Besides, as expected in Remark \ref{rem7}, $R_t^{*}$ can only act as an upper bound when the phase errors are uniformly distributed over $[-\pi, \pi)$. 

Fig.~\ref{f2_v2} illustrates the average achievable rate of user $R$ against the number of elements. The proposed upper bound is relatively loose for a small number of elements and becomes tight as the number of elements grows large. `Upper bound' in the legend stands for $\log_2(1 + q_r^2/q_t^2)$. As can be seen from Fig.~\ref{f2_v2}, the average achievable rate of user $R$ converges to $\log_2(1 + q_r^2/q_t^2)$ with increasing numbers of elements.

\subsection{Average Achievable Rates versus Transmit SNR}
\begin{figure}[!]
    \centerline{\includegraphics[scale=0.58]{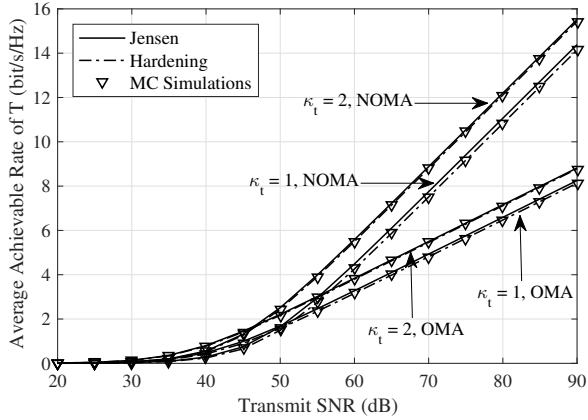}}   
    \caption{Average achievable rate of user $T$ versus transmit SNR.}
    \label{f3_v2}
\end{figure}
\begin{figure}[!]
    \centerline{\includegraphics[scale=0.58]{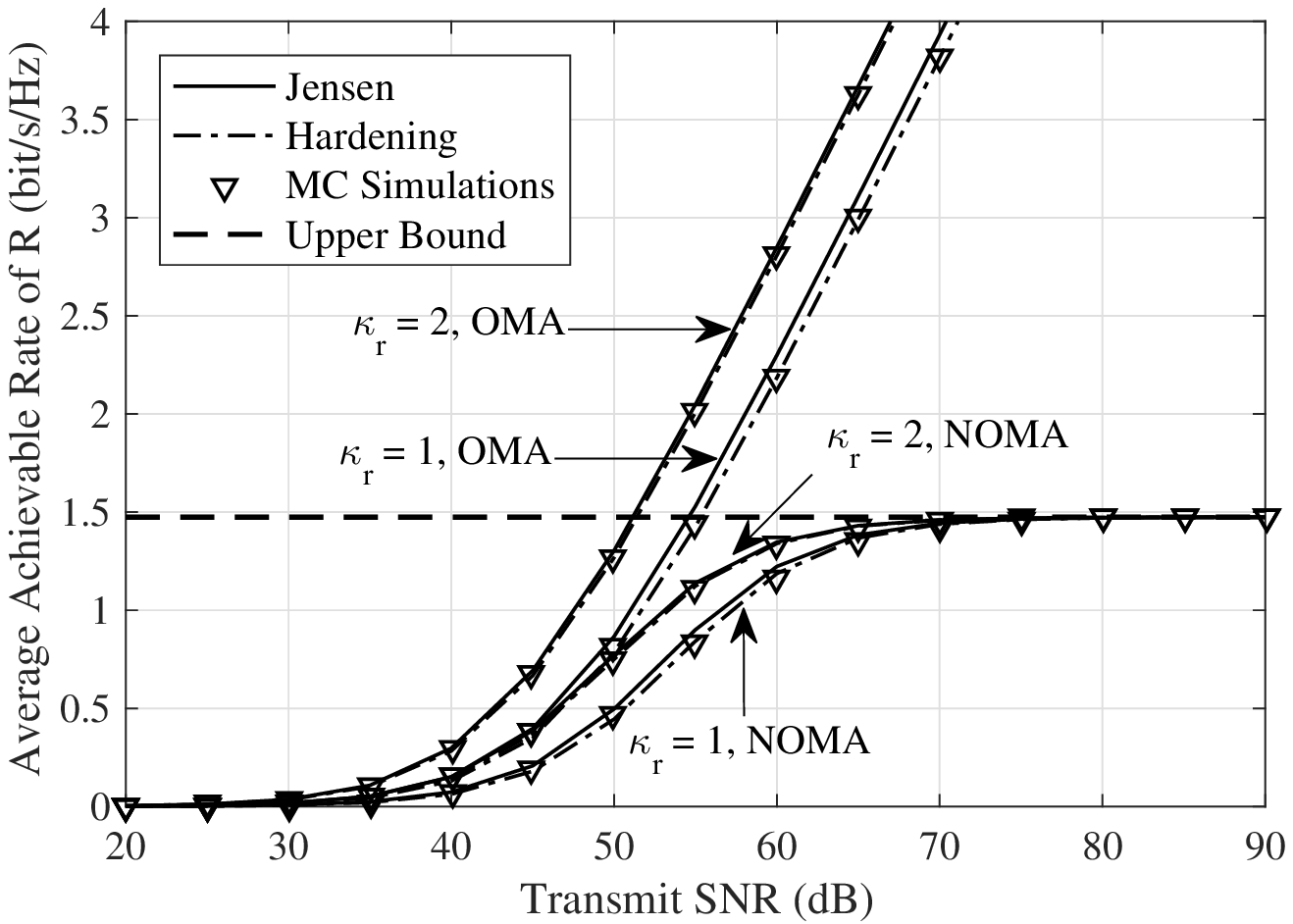}}   
    \caption{Average achievable rate of user $R$ versus transmit SNR.}
    \label{f4_v2}
\end{figure}
\begin{figure}[!]
    \centerline{\includegraphics[scale=0.6]{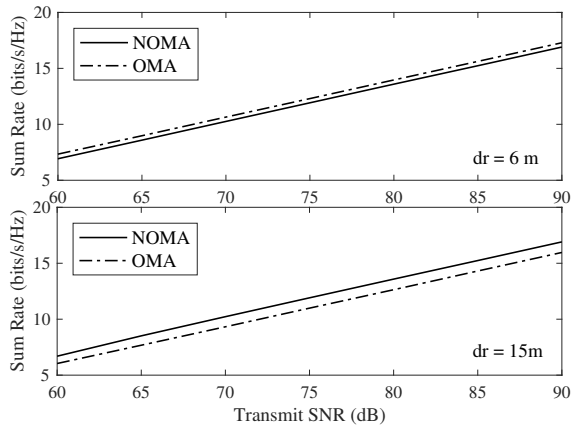}}   
    \caption{Sum rate versus transmit SNR, $\kappa_t = \kappa_r = 2$.}
    \label{f_sum}
\end{figure}
The relation between average achievable rates and transmit SNR is explored in Fig.~\ref{f3_v2}, where a rectangular IOS with $N_v = 4$ elements per column and $N_h = 10$ elements per row is used. The transmit SNR lies in the interval $\gamma_0 \in [20, 90]$ {\rm dB}. Channel estimation errors are considered by setting the concentration parameter of the Von Mises distribution to $\kappa_u \in \{1, 2\}$, $u\in\{t, r\}$. It can be observed from Fig.~\ref{f3_v2} that the average achievable rate of $T$ increases with the transmit SNR, as expected. The proposed upper bound and approximation are very close to each other. In addition, in Fig.~\ref{f4_v2}, the average achievable rate of $R$ grows and converges to the upper bound $\log_2(1 + q_r^2/q_t^2)$ with increasing SNR.  This observation corroborates our discussion in Sec. III. {  The results of OMA are also given for comparison. We can see that the average achievable rate of $T$ in NOMA outperforms that of OMA in the large transmit SNR regime. However, the average achievable rate of $R$ in NOMA is inferior to that of OMA because it converges to a constant value. As a step further, we compare the system sum rates of NOMA and OMA in Fig.~\ref{f_sum}. Two locations of $R$, i.e., $d_r = 6\,{\rm m}$ and $d_r = 15\,{\rm m}$, are considered. The large transmit SNR regime is focused. $\kappa_t = \kappa_r = 2$. For $d_r = 6\,{\rm m}$, $\alpha^4 \epsilon_t^2 \eta_t < \epsilon_r^2 \eta_r$, and OMA is better; For $d_r = 15\,{\rm m}$, $\alpha^4 \epsilon_t^2 \eta_t > \epsilon_r^2 \eta_r$, and NOMA has larger sum rate. Numerical results confirm our analysis.}

\subsection{Correlated Channels versus Uncorrelated Channels}
\begin{figure}[!]
    \centerline{\includegraphics[scale=0.58]{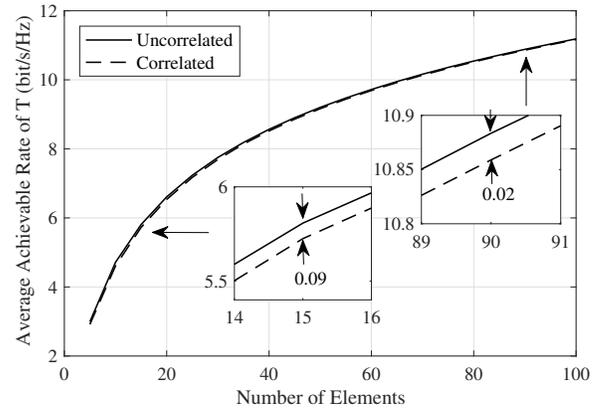}}   
    \caption{Average achievable rate of user $T$ versus transmit SNR for different channel conditions.}
    \label{f5_v2}
\end{figure}
\begin{figure}[!]
    \centerline{\includegraphics[scale=0.63]{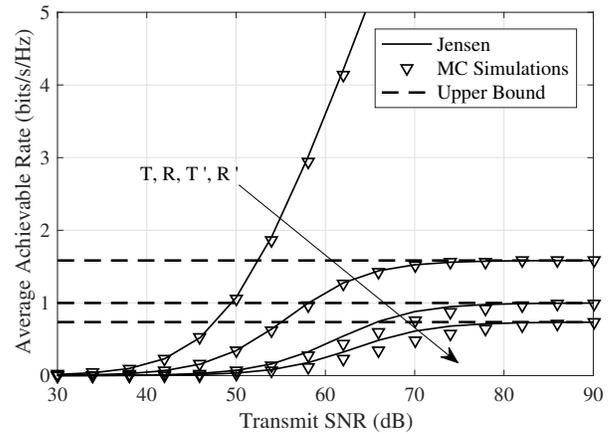}}   
    \caption{Average achievable rates of $T$, $R$, $T'$ and $R'$ versus the transmit SNR.}
    \label{fff_mul}
\end{figure}
The impact of channel correlations on the system performance is investigated in Fig.~\ref{f5_v2}, where a 1-bit phase adjustment is performed at the IOS. We consider a rectangular IOS, where each IOS element is of size $5\,{\rm cm} \times 5\,{\rm cm}$ and is placed in a $5 \times N_h$ rectangular array, with $N_h$ varying in the interval $[1, 20]$. We assume the wavelength $\lambda = 20\,{\rm cm}$. According to \eqref{corr_m_n}, the channels are correlated in this case. The average achievable rate with uncorrelated channels is provided as a benchmark\footnote{In practice, the channels associated with the IOS are uncorrelated when the elements are placed in a linear array, and the spacing of elements equals $\lambda/2$ times integers. Besides, the covariance matrix $R$  of uncorrelated channels in \eqref{corr_m_n} is the identity matrix.}. As shown in Fig.~\ref{f5_v2}, the average achievable rates of uncorrelated and correlated channels are close to each other, even for a small number of elements. For example, when $N = 15$, the performance gap is $0.09\,{\rm bit/s/Hz}$. Moreover, the performance gap of the average achievable rates between uncorrelated and correlated channels decreases with an increase in the number of elements, as expected from Proposition \ref{prop66}. For example, for $N = 90$, the performance gap decreases to $0.02\,{\rm bit/s/Hz}$. 

{ \subsection{Multiple Users NOMA}
Two more users $T'$ and $R'$ are added to the system. The distances from the IOS to $T'$ and $R'$ are $d_{t'} = 12\,{\rm m}$ and $d_{r'} = 15\,{\rm m}$, respectively. The transmit amplitude coefficients at TX are adjusted to be $q_{r'}^2 = 0.4$, $q_{t'}^2 = 0.3$, $q_{r}^2 = 0.2$ and $q_{t}^2 = 0.1$. We use a rectangular IOS with $N_v = 4$ elements per column and $N_h = 10$ elements per row and 1-bit phase adjustment is performed. Other simulation parameters are not changed. Fig.~\ref{fff_mul} shows the average achievable rates of $T$, $R$, $T'$ and $R'$ versus the transmit SNR. As expected, the average achievable rates of $R$, $T'$ and $R'$ respectively converge to $\log_2\left(1 + {q_{r}^2}/{q_t^2}\right)$, $\log_2\left(1 + {q_{t'}^2}/{(q_t^2 + q_r^2)}\right)$ and $\log_2\left(1 + {q_{r'}^2}/({q_t^2 + q_r^2 + q_{t'}^2})\right)$ as the transmit SNR goes large. In a multi-user NOMA system, the performance of users with higher decoding orders, e.g., $T'$ and $R'$ in this system, is limited and highly determined by the transmit amplitude coefficients at TX. Therefore, in existing literatures, it is more common to consider two-user NOMA, in which two paired users are grouped into the same resource block and different NOMA pairs are allocated with orthogonal resource blocks.
}

\section{Conclusions}
This paper proposed an IOS-assisted downlink NOMA network with spatially correlated channels and phase errors and analyzed the average achievable rates of two users. Specifically, we investigated an upper bound on the average achievable rates using Jensen's inequality. The channel hardening effect in the proposed system was exploited, from which an approximation on the average achievable rates was derived. The proposed upper bound and approximation were proved to be asymptotically equivalent. We also proved that the average achievable rates with correlated and uncorrelated channels are asymptotically equivalent for a large number of elements. { IOS assisted multi-user NOMA and IOS assisted OMA were also analyzed with the same framework. Numerical results illustrated that the performance gap between low-resolution elements with two-bit phase shifting and the ideal elements with continuous phase shifting is narrow and confirmed that NOMA is not always superior to OMA due to the reconfigurability of IOS over different time slots. }

\begin{appendices}

\setcounter{equation}{0}
\renewcommand\theequation{A.\arabic{equation}}
\section{}
In the proposed system model, $h_{n}$ can be written as
\begin{equation}
    h_{n} = R_n \exp\left(j \vartheta_n\right) = X_n + j Y_n,\, n \in \{1, 2,..., N\},
\end{equation}
where $R_n$, $\vartheta_n$, $X_n$ and $Y_n$ denote the amplitude, phase angle, real part and imaginary part of $h_n$, respectively; $X_n$ and $Y_n$ are uncorrelated normal random variables, each with zero mean and variance as
\begin{equation}
    \tau \triangleq {\rm Var}[X_n] = {\rm Var}[Y_n] = \frac{1}{2}.
\end{equation}
For the channels associated with two different IOS elements, i.e., $h_n$ and $h_i$, it can be obtained that
\begin{subequations}
        \begin{align}
            \mathbb{E}\left[X_n X_i\right] &= \mathbb{E}\left[|h_{n}| |h_{i}| \cos(\vartheta_n) \cos(\vartheta_i)\right], \\
            \mathbb{E}\left[Y_n Y_i\right] &= \mathbb{E}\left[|h_{n}| |h_{i}| \sin(\vartheta_n) \sin(\vartheta_i)\right] \nonumber\\
            &= \mathbb{E}\left[|h_{n}| |h_{i}| \cos(\vartheta_n + \pi/2) \cos(\vartheta_i + \pi/2)\right].
        \end{align}
\end{subequations}
Since $\vartheta_n$, $n \in\{1, 2,..., N\}$, are uniformly distributed over $[-\pi, \pi)$ and the above functions are periodic, we can have 
\begin{equation}
    \tau_1 \triangleq \mathbb{E}\left[X_n X_i\right] = \mathbb{E}\left[Y_n Y_i\right].
\end{equation}
Similarly,
\begin{subequations}
    \begin{align}
        \mathbb{E}\left[X_n Y_i\right] &= \mathbb{E}\left[|h_{n}| |h_{i}| \cos(\vartheta_n) \sin(\vartheta_i)\right], \\
        \mathbb{E}\left[Y_n X_i\right] &= \mathbb{E}\left[|h_{n}| |h_{i}|  \sin(\vartheta_n) \cos(\vartheta_i)\right] \nonumber\\
        &= -\mathbb{E}\left[|h_{n}| |h_{i}| \cos(\vartheta_n + \pi/2) \sin(\vartheta_i + \pi/2)\right].
    \end{align}
\end{subequations}
Hence,
\begin{equation}
    \tau_2 \triangleq \mathbb{E}\left[X_n Y_i\right] = - \mathbb{E}\left[Y_n X_i\right].
\end{equation}
For $n \neq i$, the joint distribution of $(X_n, Y_n, X_i, Y_i)$ can be written as
\begin{equation}
    f(x_n, y_n, x_i, y_i) = \frac{1}{(2 \pi)^2\sqrt{{\rm det} \mathbf{C}}} \exp\left(-\frac{1}{2} \mathbf{v} \mathbf{C}^{-1} \mathbf{v}^{T}\right),
    \label{jpdf}
\end{equation}
where $\mathbf{v} = [x_n, y_n, x_i, y_i]$; $\mathbf{C}$ is the covariance matrix of $\{X_n, Y_n, X_i, Y_i\}$, given by 
\begin{equation}
    \mathbf{C} = 
    \begin{pmatrix}
        \tau   &   0     & \tau_1   & \tau_2 \\
        0      & \tau    & -\tau_2  & \tau_1\\
        \tau_1 & -\tau_2 & \tau     & 0\\
        \tau_2 & \tau_1  & 0        & \tau
    \end{pmatrix}.
\end{equation}
The determinant and inverse of $\mathbf{C}$ can be calculated as
\begin{equation}
    {\rm det} \{\mathbf{C}\} = (\tau^2 - \tau_1^2 - \tau_2^2)^2,
    \label{det_c}
\end{equation}
\begin{equation}
\mathbf{C}^{-1} = 
    \begin{pmatrix}
        \frac{\tau}{\tau^2 - \tau_1^2 - \tau_2^2}  & 0 & \frac{-\tau_1}{\tau^2 - \tau_1^2 - \tau_2^2} & \frac{-\tau_2}{\tau^2 - \tau_1^2 - \tau_2^2} \\[5pt]
        0 & \frac{\tau}{\tau^2 - \tau_1^2 - \tau_2^2} & \frac{\tau_2}{\tau^2 - \tau_1^2 - \tau_2^2} & \frac{-\tau_1}{\tau^2 - \tau_1^2 - \tau_2^2} \\[5pt]
        \frac{-\tau_1}{\tau^2 - \tau_1^2 - \tau_2^2} & \frac{\tau_2}{\tau^2 - \tau_1^2 - \tau_2^2} & \frac{\tau}{\tau^2 - \tau_1^2 - \tau_2^2} & 0 \\[5pt]
        \frac{-\tau_2}{\tau^2 - \tau_1^2 - \tau_2^2} & \frac{-\tau_1}{\tau^2 - \tau_1^2 - \tau_2^2} & 0 & \frac{\tau}{\tau^2 - \tau_1^2 - \tau_2^2}
    \end{pmatrix}.
    \label{inv_c}
\end{equation}
Taking \eqref{det_c} and \eqref{inv_c} into \eqref{jpdf}, the joint distribution of $(X_n, Y_n, X_i, Y_i)$ can be obtained. Then, using the theorem of the transformation of random variables:
\begin{equation}
    f(r_n, r_i, \vartheta_n, \vartheta_i) = \left| {\rm det}(J(r_n, r_i, \vartheta_n, \vartheta_i)) \right | f(x_n, y_n, x_i, y_i),
\end{equation}
where 
\begin{equation}
    J(r_n, r_i, \vartheta_n, \vartheta_i) = \frac{\partial(x_n, y_n, x_i, y_i)}{\partial(r_n, r_i, \vartheta_n, \vartheta_i)}
\end{equation} 
is  the  Jacobian  matrix  for the transformation
\begin{equation}
    \begin{split}
        \{&x_n = r_n \cos(\vartheta_n), y_n = r_n \sin(\vartheta_n), \\
        &x_i = r_i \cos(\vartheta_i), y_i = r_i \sin(\vartheta_i)\},
    \end{split}
\end{equation}
we can derive the joint distribution of $(R_n, R_i, \vartheta_n, \vartheta_i)$ as
\begin{equation}
    \begin{split}
        &f(r_n, r_i, \vartheta_n, \vartheta_i) = \frac{r_n r_i}{(2\pi)^2 (\tau^2 - q^2)} \\
        &~~\times \exp\left[ - \frac{\tau\left(r_n^2 + r_i^2\right) - 2 r_n r_i q \cos\left(\vartheta_i - \vartheta_n - \xi\right)}{2 \left(\tau^2 - {q^2} \right)}\right],
    \end{split}
\end{equation}
where 
\begin{subequations}
    \begin{align}
        \xi &= \arctan\left(\frac{\tau_2}{\tau_1}\right),\\
        q^2 &= \tau_1^2 + \tau_2^2 = \frac{\left|\mathbb{E}[h_n h_i^{*}]\right|^2}{4}.
    \end{align}
\end{subequations}
Using \cite[eq.6.631.1, eq.7.621.4]{integral} and after some mathematical manipulation, $\mathbb{E}[R_n R_i]$ can be expressed as
\begin{equation}
    \begin{split}
        &\mathbb{E}[R_n R_i] \\
        =&\int_0^{\infty} \int_0^{\infty} \int_0^{2 \pi} \int_0^{2 \pi} r_n r_i f(r_n, r_i, \vartheta_n, \vartheta_i) \rmd \vartheta_n \rmd \vartheta_i \rmd r_n \rmd r_i \\[5pt]
        =& \left(\frac{q^2}{\tau}-\tau\right) K\left(\frac{q^2}{\tau^2}\right)+2 \tau E\left(\frac{q^2}{\tau^2}\right).
    \end{split}
\end{equation}
For $n = i$, it is easy to show that
\begin{equation}
    \mathbb{E}[R_n R_i] = 2 \tau = 1.
\end{equation}

The same conclusions can be derived for the channels $\mathbf{g}$ and $\mathbf{r}$ following the framework. Thus, the statement in Proposition \ref{prop11} is proved.

\setcounter{equation}{0}
\renewcommand\theequation{B.\arabic{equation}}
\section{}
Define
\begin{equation}
    f(x) = \left(\frac{x}{2} - \frac{1}{2}\right) K\left({x}\right) +  E\left({x}\right).
\end{equation}
It can be proved that $f(x)$ is an increasing function for $x \geq 0$. Since 
\begin{equation}
    0 \leq \left|\mathbb{E}[w_n w_i^{*}]\right|^2 \leq 1,
    \label{85_exp_ran}
\end{equation}
taking \eqref{85_exp_ran} into \eqref{exp_e_h_h}, we can have
\begin{equation}
    \frac{\pi}{4} \leq \mathbb{E}\left[\left|w_n\right|\left|w_i\right|\right] \leq 1.
\end{equation}

\setcounter{equation}{0}
\renewcommand\theequation{C.\arabic{equation}}
\section{}
Since 
\begin{equation}
    \mathbb{E}\left[\sum\limits_{n=1}^{N} \left|g_{n}\right|^2 \left|h_n\right|^2 \right] = N,
    \label{sum_square1}
\end{equation}
we can have
\begin{equation} \small
    \mathbb{E}\left[\sum\limits_{n=1}^{N-1}\sum\limits_{i=n+1}^{N} \left|g_{n}\right|\left|h_n\right| \left|g_{i}\right|\left|h_i\right| \right] = \frac{\mathbf{tr}\left(\mathbf{\Bar{R} \Bar{R}}\right) - N}{2}.
\end{equation}
Therefore,
\begin{equation}
    \begin{split}
        &\mathbb{E}\left[\sum\limits_{n=1}^{N-1}\sum\limits_{i=n+1}^{N} \left|g_{n}\right|\left|h_n\right| \cos(\phi_{n}^{t}) \left|g_{i}\right|\left|h_i\right| \cos(\phi_{i}^{t})\right] \\[5pt]
        &= \frac{\epsilon_t^2}{2} \left(\mathbf{tr}\left(\mathbf{\Bar{R} \Bar{R}}\right) - N\right).
        \label{sum_cos}
    \end{split}
\end{equation}
Moreover,
\begin{equation}
    \mathbb{E}\left[\sum\limits_{n=1}^{N-1}\sum\limits_{i=n+1}^{N} \left|g_{n}\right|\left|h_n\right| \sin(\phi_{n}^{t}) \left|g_{i}\right|\left|h_i\right| \sin(\phi_{i}^{t})\right] = 0,
    \label{sum_sin}
\end{equation}
due to the symmetrical distribution around zero of $\phi_{n}^{t}$, $n \in \{1,..., N\}$. Substituting \eqref{sum_square1}, \eqref{sum_cos} and \eqref{sum_sin} into \eqref{decom_t}, we can obtain \eqref{r_t_*}. Thus, the proof is complete.

\setcounter{equation}{0}
\renewcommand\theequation{D.\arabic{equation}}
\section{}
To begin with, a Lemma regarding the sum of correlated random variables is presented as follows.
\begin{lemma}[\cite{cacoullos2012exercises}]
    Consider a sequence of random variables $\{W_k\}$. If $\{W_k\}$ satisfies
    \begin{itemize}
        \item $\mathbb{E}\left[W_k\right] = a$,
        \item ${\rm Var}\left[W_k\right]$ is bounded,
        \item ${\rm Cov}\{W_i, W_j\} \to 0$ as $|i-j| \to \infty$,
    \end{itemize}
    then, 
    \begin{equation}
        \frac{1}{K} \sum\limits_{k=1}^{K} W_k \overset{P}{\to} a,\,\, K \to \infty,
    \end{equation}
    with the convergence in probability.
    \label{lemm2}
\end{lemma}
\begin{IEEEproof}
    Even though this is a known result, we include a proof because the obtained scaling of the variance with $K$ is used in this paper in the proof of Proposition \ref{prop4}. 
    
    Since the variance is bounded, the covariance of $W_i$ and $W_j$ is bounded as
    \begin{equation}
        {\rm Cov}\left\{W_i, W_j\right\} \leq c_0,
    \end{equation}
    where $c_0$ is a constant. Denoting $V_K = \sum\limits_{k=1}^{K} W_k$, the variance of $V_K$ can be written as
    \begin{align}
        {\rm var}[V_K] &= \sum\limits_{k=1}^{K} {\rm Cov}\left\{W_k, W_k\right\} + 2 \sum\limits_{i=1}^{K-1} \sum\limits_{j=i+1}^{K} {\rm Cov}\left\{W_i, W_j\right\}\nonumber\\
        &\leq K c_0 + 2 \sum\limits_{i=1}^{K-1} \sum\limits_{j=i+1}^{K} {\rm Cov}\left\{W_i, W_j\right\}.
    \end{align}
    Since ${\rm Cov}\left\{W_i, W_j\right\} \to 0$ when $|i-j| \to \infty$, we can find $K_0$ such that for $|i-j| > K_0$, ${\rm Cov}\left\{W_i, W_j\right\} < \varepsilon$, for any $\varepsilon > 0$. Hence, 
    {\small\begin{align}
        &\left| \sum\limits_{i=1}^{K-1} \sum\limits_{j=i+1}^{K} {\rm Cov}\left\{W_i, W_j\right\}\right | \nonumber\\
         \leq& \left| \sum\limits_{i=1}^{K-1} \sum\limits_{j=i+1}^{i + K_0} {\rm Cov}\left\{W_i, W_j\right\}\right |
        + \left| \sum\limits_{i=1}^{K-1} \sum\limits_{j=i + K_0+1}^{K} {\rm Cov}\left\{W_i, W_j\right\}\right | \nonumber\\
         \leq& K K_0 c_0 + K^2 \varepsilon.
    \end{align}}
    Therefore, the variance of $V_K/K$ satisfies 
    \begin{equation}
        {\rm Var}\left[\frac{V_K}{K}\right] \leq \frac{c_0}{K} + \frac{2 K_0 c_0}{K} + 2 \varepsilon.
    \end{equation}
    It can be seen that ${\rm Var}\left[\frac{V_K}{K}\right] \to 0$, as $K \to \infty$. Thus, the proof of Lemma \ref{lemm2} is complete.
\end{IEEEproof}

The composite channel associated with user $T$ is analyzed first. Let $Y_n = |g_n||h_n|e^{j\phi_n^t}$. We can have
\begin{equation}
    \mathbb{E}[Y_n] = \frac{\pi}{4} \epsilon_t,
\end{equation}
and the variance of $Y_n$ is bounded. Covariance of $\{Y_n, Y_i\}$ can be written as
\begin{equation}
    \begin{split}
        {\rm Cov}\{Y_n, Y_i\} =& \mathbb{E}\left[|g_n||h_n|e^{j\phi_n^t}|g_i||h_i|e^{-j\phi_i^t}\right] \\
        &- \mathbb{E}\left[|g_n||h_n|e^{j\phi_n^t}\right] \mathbb{E}\left[|g_i||h_i|e^{-j\phi_i^t}\right].    
    \end{split}
\end{equation}
From \eqref{corr_m_n} and \eqref{exp_e_h_h}, it can be learned that 
\begin{equation}
    \mathbb{E}\left[|w_n||w_i|\right] \to \mathbb{E}\left[|w_n|\right] \mathbb{E}\left[|w_i|\right] = \frac{\pi}{4},\,\, |n-i| \to \infty,
\end{equation}
where $w \in \{h, g, r\}$.
Given that the phase errors are i.i.d. with symmetric distribution around zero, it can be proved that
\begin{equation}
    \begin{split}
        {\rm Cov}\{Y_n, Y_i\} =&\,\mathbb{E}\left[|g_n||g_i|\right] \mathbb{E}\left[|h_n||h_i|\right]  \epsilon_t^2 \\ 
        &- \mathbb{E}\left[|g_n|\right] \mathbb{E}\left[|g_i|\right] \mathbb{E}\left[|h_n|\right] \mathbb{E}\left[|h_i|\right] \epsilon_t^2\\ 
    \to&\,0,\,\, |n-i| \to \infty.
    \end{split}
\end{equation}
Hence, using Lemma \ref{lemm2}, we obtain 
\begin{equation}
    \frac{1}{N} \sum\limits_{n=1}^{N}\left|g_{n}\right| \left|h_n\right| e^{j\phi_{n}^{t}} \overset{P}{\to} \frac{\pi}{4} \epsilon_t,\,\, N \to \infty.
    \label{g_h_asy}
\end{equation}
With the continuous mapping theorem, \eqref{h_t_n^2} in Proposition \ref{prop33} can be obtained. Following the same framework, \eqref{h_r_n^2} can be derived. 

\setcounter{equation}{0}
\renewcommand\theequation{E.\arabic{equation}}
\section{}
Let 
\begin{subequations}
    \begin{align}
        A_N &= \sum\limits_{n=1}^{N}\left|g_{n}\right| \left|h_n\right| \cos(\phi_n^t), \\
        B_N &= \sum\limits_{n=1}^{N}\left|g_{n}\right| \left|h_n\right| \sin(\phi_n^t).
    \end{align}
\end{subequations}
The expectations of $A_N$ and $B_N$ are $\mathbb{E}[A_N] = \frac{\pi}{4} N \epsilon_t$ and $\mathbb{E}[B_N] = 0$. $\mathbb{E}[A_N^2 + B_N^2]$ can be written as
\begin{equation}
    \mathbb{E}[A_N^2 + B_N^2] = \mathbb{E}^2[A_N] + {\rm var}[A_N] + \mathbb{E}^2[B_N] + {\rm var}[B_N].
\end{equation}
As shown in Lemma \ref{lemm2}, the variances of $A_N$ and $B_N$ scale with $N$. Thus, when the phase errors are non-uniform over $[-\pi, \pi)$, $\mathbb{E}[A_N^2 + B_N^2]$ satisfies the asymptotic equivalence as
\begin{equation}
    \mathbb{E}[A_N^2 + B_N^2] \sim \frac{\pi^2}{16} N^2 \epsilon_t^2,~~~ N \to \infty.
    \label{ap_e_99}
\end{equation}
Thus, with the properties of asymptotic analysis, we can obtain \eqref{r_t*4a} in Proposition \ref{prop4}. Following the similar procedures, \eqref{r_t*4b} can be derived as well. 


\setcounter{equation}{0}
\renewcommand\theequation{F.\arabic{equation}}
\section{}
As shown in Proposition \ref{prop33}, the random variable $G_N = \frac{1}{N^2} H_t$ satisfies 
\begin{equation}
    G_N \overset{P}{\to} \frac{\pi^2}{16} \epsilon_t^2,~~~ N \to \infty,
\end{equation}
with the convergence in probability. As shown in Proposition \ref{prop4} and its proof, the expectation and variance of $G_N$ satisfy
\begin{equation}
    \mathbb{E}[G_N] \to \frac{\pi^2}{16} \epsilon_t^2,~~ {\rm Var}[G_N] \to 0,~~~ N \to \infty.
\end{equation}
Thus, when the phase errors are non-uniform over $[-\pi, \pi)$, the expectation and variance of $H_t$ satisfy
\begin{equation}
    \frac{{\rm Var}\left[H_t\right]}{\mathbb{E}^2\left[H_t\right]} \to 0,~~~ N \to \infty.
    \label{asy_con}
\end{equation}
Invoking \cite[Theorem 4]{sanayei2007opportunistic}, the asymptotic equivalence of $R_t$ and $R_t^{*}$ can be obtained. 
With the transitive property of asymptotic analysis, the asymptotic equivalence of $R_t$ and $R_t^{+}$ also holds. 

When the phase errors are uniform over $[-\pi, \pi)$, the expectation of $G_N$ goes to zero as $N$ goes large. The asymptotic equivalence of $R_t$ and $R_t^{*}$ does not hold since the relation in \eqref{asy_con} is not satisfied.

{ \setcounter{equation}{0}
\renewcommand\theequation{G.\arabic{equation}}
\section{}
$R_{r'}$ is analyzed first. $\gamma_{r'}$, $\gamma_{t'\to r'}$, $\gamma_{r\to r'}$ and $\gamma_{t\to r'}$ are given by
\begin{subequations}
    \begin{align}
        \gamma_{r'} &= \frac{P \eta_{r'} \beta^2 H_{r'} q_{r'}^2}{P \eta_{r'} \beta^2 H_{r'} \left(q_t^2 + q_r^2 + q_{t'}^2\right) + \sigma_0^2}, \\
        \gamma_{t'\to r'} &= \frac{P \eta_{t'} \alpha^2 H_{t'} q_{r'}^2}{P \eta_{t'} \alpha^2 H_{t'} \left(q_t^2 + q_r^2 + q_{t'}^2\right) + \sigma_0^2},\\
        \gamma_{r\to r'} &= \frac{P \eta_{r} \beta^2 H_{r} q_{r'}^2}{P \eta_{r} \beta^2 H_{r} \left(q_t^2 + q_r^2 + q_{t'}^2\right) + \sigma_0^2}\\
        \gamma_{t\to r'} &= \frac{P \eta_{t} \alpha^2 H_{t} q_{r'}^2}{P \eta_{t} \alpha^2 H_{t} \left(q_t^2 + q_r^2 + q_{t'}^2\right) + \sigma_0^2},
        \end{align}
\end{subequations}
respectively, where $H_{t}$ and $H_{r}$ is defined in \eqref{hthr}. $H_{t'}$ and $H_{r'}$ can be written as
\begin{equation}
    H_{t'} = \left|\sum\limits_{n=1}^{N}\left|g'_{n}\right| \left|h_n\right| e^{j\phi_{n}^{t'}}\right|^2, H_{r'} = \left|\sum\limits_{n=1}^{N}\left|r'_{n}\right| \left|h_n\right| e^{j\phi_{n}^{r'}}\right|^2.
\end{equation}
where $g'_{n}$ and $r'_{n}$ are the channels from the $n$th element to $T'$ and $R'$, respectively. Since the priority oriented scheme is used and phase shifts are adjusted to boost $H_t$ and $H_r$, $\phi_{n}^{t'}$ and $\phi_{n}^{r'}$ can be regarded as uniformly distributed random variables over $[-\pi, \pi)$. Following the similar procedures in Propositions \ref{prop11} and \ref{prop22}, an upper bound of $R_{r'}$ is derived by
\begin{equation}
    R_{r'}^{*} = \underset{\zeta \in\{t,r,t',r'\}}{\min}\left\{ \log_2\left(1 + \frac{q_{r'}^2}{q_t^2 + q_r^2 + q_{t'}^2 + f_{\zeta}^{-1} }\right)\right\},
\end{equation}
Since $f_{t'}<f_{t}$ and $f_{r'}<f_{r}$, $R_{r'}^{*}$ can be simplified as in \eqref{r_r_4_*}. Following the same method, $R_{t'}^{*}$ can be also derived.
}

\end{appendices}

\bibliographystyle{IEEEtran}
\bibliography{irs}

\end{document}